\documentclass[aps,prd,twocolumn,superscriptaddress,amsfonts,amssymb,amsmath,eqsecnum,nofootinbib,floatfix]{revtex4}
\usepackage{graphicx}
\usepackage{bm}
\usepackage{dcolumn}

\begin{document}

\title{Reducing Thermoelastic Noise in Gravitational-Wave Interferometers by Flattening the Light Beams}

\author{Erika D'Ambrosio}
\affiliation{LIGO Laboratory, California Institute of Technology, Pasadena, CA 91125}

\author{Richard O'Shaughnessy}
\affiliation{Theoretical Astrophysics, California Institute of Technology, Pasadena, CA 91125}

\author{Sergey Strigin}
\affiliation{Physics Faculty, Moscow State University, Moscow, Russia 119992}

\author{Kip S. Thorne}
\affiliation{Theoretical Astrophysics, California Institute of Technology, Pasadena, CA 91125}

\author{Sergey Vyatchanin}
\affiliation{Physics Faculty, Moscow State University, Moscow, Russia 119992}

\date{Received 13 September 2004}

\begin{abstract}

In the baseline design for advanced LIGO interferometers, the most
serious noise source is tiny, dynamically fluctuating bumps and 
valleys on the faces of the arm-cavity mirrors, caused by random flow of heat in the mirrors' sapphire substrates: 
so-called {\it thermoelastic noise}. 
We propose replacing the interferometers' baseline arm-cavity light beams,
which have {\it Gaussian}-shaped intensity profiles that do not average 
very well over the dynamical bumps and valleys, by beams with 
{\it mesa}-shaped profiles that are flat in their central $\sim 7$ cm of
radius, and that then fall toward zero as quickly as is allowed by
diffraction in LIGO's 4 km arms; see Fig.\ \ref{fig:MHGaussModes}.  The mesa beams average the bumps and
valleys much more effectively than the Gaussian beams. As a result, 
if the mirrors' substrate radii and thicknesses are held fixed at 15.7 cm 
and 13 cm, and the beam radii are adjusted so diffraction losses
per bounce are about 10 ppm, replacing Gaussian beams by mesa beams reduces the 
thermoelastic noise power by about a factor 3.  If other thermal noises 
are kept negligible, this reduction will permit advanced LIGO to
beat the Standard Quantum Limit (circumvent the Heisenberg Uncertainty
Principle for 40 kg mirrors) by about a factor 1.5 over a bandwidth about equal to
frequency; optical (unified quantum) noise will become the dominant
noise source; and the event rate for inspiraling neutron star binaries
will increase by about a factor 2.5.  
The desired mesa beams can be produced from input,
Gaussian-profile laser light, by changing the shapes of the arm
cavities' mirror faces from their baseline {\it spherical} shapes (with
radii of curvature of order 60 km) to {\it Mexican-Hat} ({\it MH};
sombrero-like)
shapes that have a shallow bump in the center but are otherwise much flatter
in the central 10 cm than
the spherical mirrors, and then flare upward
strongly in the outer 6 cm, like a sombrero; Fig.\  \ref{fig:MHGaussMirrors}.
In this paper we describe mesa beams and MH mirrors mathematically and
we report the results of extensive 
modeling calculations, which show that the mesa-beam interferometers are 
{\it not} substantially more sensitive than the baseline Gaussian-beam
interferometers to errors in the mirror figures, positions, and
orientations.  
This has motivated the LIGO Scientific Community (LSC) to 
adopt MH mirrors and mesa beams as an option for advanced LIGO, to be studied
further.  The details of our modeling calculations are presented in companion papers.

\end{abstract}

\pacs{04.80.Nn,95.55.Ym,07.60.Ly}

\maketitle

\section{Introduction and Summary \label{sec:Intro} }

The Laser Interferometer Gravitational-Wave Observatory (LIGO) is designed to
support successive generations of interferometric gravitational-wave detectors.
LIGO's first interferometers are now in operation \cite{LSC0}, and the (negative) results of
its first gravitational-wave searches have recently been 
published.
\cite{LSCn}.  When they reach their design sensitivity (presumably next year), LIGO's
initial interferometers, together with their international partners, will reach out into
the universe to distances where it is plausible, but not probable to detect gravitational waves
\cite{CutlerThorne}.  After a planned upgrade to {\it advanced LIGO interferometers}
(planned to begin in 2007), wave detection will be quite probable \cite{CutlerThorne}.
A baseline design for the advanced LIGO interferometers has recently been
adopted \cite{ALDesign}, along with several options, not currently in the baseline,
that merit further study and might be incorporated at a future date.  This paper
describes one of these options, which has been much discussed within the LIGO
Scientific Community (LSC) but has not previously been presented in the published
literature: the reshaping of the arm-cavity light beams so as to reduce thermoelastic
noise.

\subsection{The Context: Noise in Advanced LIGO Interferometers \label{sec:ALNoise} }

For advanced LIGO's baseline design \cite{ALDesign}, 
the dominant noise sources in
the most interesting frequency range (above about 20 Hz) are
thermoelastic noise and optical noise (also called 
``unified quantum noise'').  Other thermal noises (most especially
coating thermal noise) might, in the end, be important; but in this
paper we shall assume them negligible and shall focus on the
thermoelastic noise and optical noise.

\begin{figure}
\includegraphics[width=3.5in]{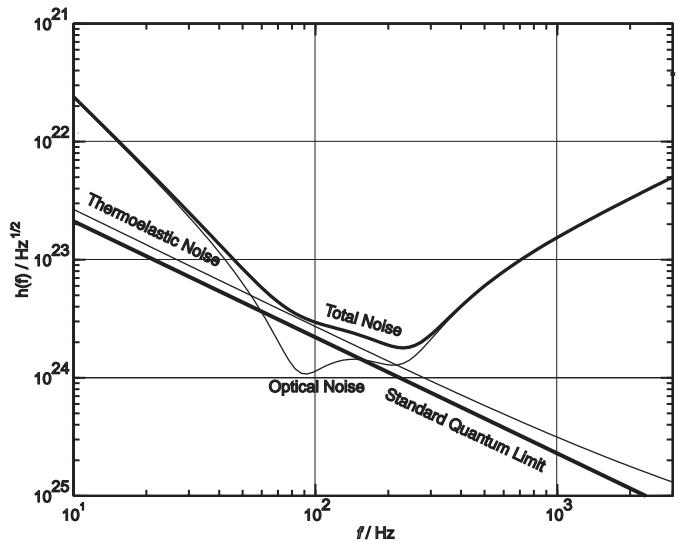}%
\caption{Noise curves for advanced LIGO.\label{fig:NoiseCurves}}
\end{figure}

In Fig.\ \ref{fig:NoiseCurves} we show the thermoelastic noise 
\cite{BGV,LiuThorne}, the
optical noise \cite{BuonannoChen1,BuonannoChen2,BuonannoChen3}, 
their sum (labeled total noise), and the standard quantum limit (SQL) for 
the advanced LIGO baseline design with sapphire mirrors \cite{ALDesign}.  
This figure suggests
(as is well known \cite{BuonannoChen1}) that, if the thermoelastic
noise can be reduced significantly (and if other thermal noises
can be kept negligible), then the advanced LIGO interferometers
will be able to beat the SQL, and the interferometers' ranges 
(detectable distances) for astrophysical sources will be increased
significantly.

In this paper we propose a method (``flattening the interferometers'
light beams'') for reducing the thermoelastic
noise, we evaluate the resulting increased range for
neutron-star / neutron-star (NS/NS) binaries, and we explore practical
issues related to our proposal.  We have previously discussed
our proposal, the increased NS/NS range, and the practical issues in
presentations at meetings of the LIGO Scientific Collaboration \cite{DOTlsc0,DOSTVlsc1,DOSTVlsc2,DOSTVmit} and in an internal LIGO document \cite{DOSTVshortdoc}.  

It is worth emphasizing that our proposed beam flattening need not entail any
major modifications of the advanced-LIGO interferometer design; it only involves
reshaping the reflecting surfaces of the test mass mirrors, adjusting the input
beam, and modifying the mirror-alignment control system to account for the changes
of the light-mode shapes.

\subsection{The Physical Nature of Thermoelastic Noise; Motivation for Reshaping Beams}

Our proposal is motivated by the physical nature of thermoelastic noise.  This noise
is created by the stochastic flow of heat (random motions of thermal phonons) within
each test mass (mirror), which produces stochastically fluctuating hot spots and
cold spots inside the test mass.  The test-mass material (sapphire for the baseline
design of advanced LIGO) expands in the hot spots
and contracts in the cold spots, creating fluctuating bumps and valleys on the test-mass
(mirror) faces.  
These face bumps influence the light beam's measurement of the test masses' 
positions:  the interferometer's output phase shift is proportional to the difference of the
test masses' average positions --- with the average being the position of a mirrored
test-mass face, weighted by the light's energy flux (its intensity distribution).

If the intensity distribution is ``flat'' (nearly constant) in most regions of high intensity, then the
adjacent valleys and bumps (having been created by heat flow from one to the other) will
average out, giving low net thermoelastic noise.  If, instead, the energy flux is sharply changing
in most regions of high flux, then the adjacent valleys and bumps will not average well and the
thermoelastic noise will be high.  Also, the larger is the light beam, the better will
be the averaging and thus the lower will be the noise.

These considerations suggest that large-radius, flat-topped beams with 
steep edges (e.g.\ the thick curve in Fig.\ \ref{fig:MHGaussModes} below)
will lead to much smaller
thermoelastic noise than small-radius, centrally peaked beams with gradually sloping sides (e.g., the thin, Gaussian curve in Fig.\ \ref{fig:MHGaussModes} 
below).

For Gaussian beams, the influence of beam radius $r_o$ has been quantified by Braginsky, Gorodetsky and Vyatchanin
\cite{BGV} (who first pointed out the importance of thermoelastic noise for sapphire test
masses): the thermoelastic noise power scales as $S_h^{\rm TE} \propto 1/r_o^3$
(aside from small corrections due to the test masses' finite sizes \cite{LiuThorne}).
This has motivated the baseline design for advanced LIGO interferometers with sapphire
test masses: the beam radius $r_o$ is chosen as large as possible, given the demand for
small diffraction losses, ${\cal L}_0 \alt 10$ ppm per bounce in the interferometer's arm
cavities.\footnote{The 10 ppm is dictated 
by the
following considerations:  For the baseline design there is 125 W of input power to the interferometer and
830 kW of circulating power in each arm cavity.  Ten ppm of diffraction loss per bounce
results in a diffraction power loss in the arm cavities of $4 \times 10 \hbox{ppm} \times 830 {\rm kW} = 33$ W, which is 25 per cent of the 125 W of input light, a reasonable value.}

The baseline design uses light beams with a Gaussian distribution of energy flux, since
such beams are eigenfunctions of cavities with spherical mirrors, and spherical mirrors
are a standard, well-developed technology.  However, the Gaussian energy flux is far
from flat: most of the energy is in regions where the flux is rapidly varying with radius (thin  curve in Fig.\ \ref{fig:MHGaussModes} below),
and correspondingly the thermoelastic noise is substantially larger than it would be with
``flat-topped'' beams (thick curve in Fig.\ \ref{fig:MHGaussModes}).  This 
has motivated a (previously unpublished) proposal by O'Shaughnessy and Thorne
\cite{DOTlsc0} to replace the Gaussian beams
with flat-topped beams, while keeping the beam radius as large as is compatible with
diffraction losses ${\cal L}_0 \alt 10$ ppm.

\subsection{Summary of Analysis and Results}
\label{sec:SummaryResults}

In Sec.\ \ref{sec:MH} we construct an example of a flat-topped light beam --- a flat-topped TEM00
mode of light that will resonate in an interferometer's arm cavity, if the test-mass
mirror faces are shaped appropriately.  Because the intensity distribution of our
flat-topped beam resembles a mesa in the southwest American desert, 
we call it a {\it mesa beam}\footnote{It is also called a {\it flat-topped beam} and a
{\it mexican-hat} or {\it MH} beam in the internal LIGO literature
\cite{DOTlsc0,DOSTVlsc1,DOSTVlsc2,DOSTVmit,DOSTVshortdoc} }
(a name suggested to us by Phil Willems).  
To produce this mesa beam as an eigenmode of a symmetric arm cavity one must give
the mirror faces a shape, with a central bump and an upturned brim, that 
resembles 
a Mexican hat (or sombrero) (Fig.\ \ref{fig:MHGaussMirrors}), so we call the mirrors 
{\it Mexican-hat (MH) mirrors}.

We have not
optimized our mesa beams' intensity distribution so as to
bring the thermoelastic noise to the lowest value possible, but in Sec.\ \ref{sec:MH}
we argue that our chosen mesa beams are likely to be close to optimal.

Figure \ref{fig:MHGaussModes} below shows the
intensity distribution for our proposed mesa beam (thick curve) 
and compares it with
the intensity distribution of the baseline Gaussian beam 
(thin curve), which has the same diffraction losses.
Figure \ref{fig:MHGaussMirrors} below compares the mirror shapes that support these
mesa and Gaussian beams as eigenmodes of a 4 km LIGO arm.  In their
inner 10 centimeters of radius, the MH mirrors that support mesa beams 
are much flatter than the spherical mirrors that support Gaussian beams, 
but in their outer 6 centimeters (the upturned brim region),
the MH mirrors are far more curved. 

As we shall see, this greater curvature at large radii 
compensates considerably for the flatter shape at small radii, enabling a
mesa-beam interferometer to exhibit only modestly worse parasitic-mode behavior than
a Gaussian-beam interferometer, and only modestly worse 
sensitivity to mirror tilts, displacements, and figure errors.

Three of us (O'Shaughnessy, Strigin and Vyatchanin; OSV \cite{OSV}) have computed
the substantial reductions in thermoelastic noise that can be achieved in advanced LIGO
by replacing the baseline
spherical mirrors and their Gaussian beams with MH mirrors and their mesa beams.
The method of computation and the results are described in Sec.\ 
\ref{sec:TENoise}.  Our principal conclusion is this: 
\begin{enumerate}
\item 
By switching from the baseline (BL) spherical mirrors 
to MH mirrors with the same cylindrical test-mass diameters 
and thicknesses and the same 10-ppm-per-bounce 
diffraction losses as the BL, one can reduce the
power spectral density of thermoelastic noise by a factor 0.34 and increase
the 
volume of the universe reachable  for compact-binary inspirals,
and thus also LIGO's binary inspiral event rate,
by a factor 2.6.  
Larger improvements
could be achieved by using conical test masses with enlarged inner faces
(frustums; Tables III and IV and Appendix I of \cite{OSV}).
\end{enumerate}

One might worry that the greater flatness of the MH mirrors, in the inner 10 cm where most
of the light resides, will make mesa-beam interferometers much more sensitive  to errors in the orientations, positions, and figures
of the mirrors.  We have explored this issue in great depth, with the conclusion that
{\it mesa-beam interferometers are {\bf not} substantially more sensitive to mirror errors
than Gaussian-beam interferometers.}  Details of 
our explorations are given in companion papers by D'Ambrosio \cite{dambrosio} and by
O'Shaughnessy, Strigin and Vyatchanin \cite{OSV}, and our methods and conclusions are 
presented and discussed in
Sec.\ \ref{sec:Practical} of this paper.   Our quantitative conclusions
rely on a comparison of two {\it fiducial} advanced-LIGO interferometer configurations, both of
which have the same $L=4$ km arm lengths, and the same $R=16$ cm coated mirror radii,
and beam radii that give the same 18ppm diffraction losses off the mirrors.
One configuration uses {\it fiducial spherical mirrors} with radii of curvature ${\mathcal R}_c
= 83$ km and so has $g=1-L/{\mathcal R}_c = 0.952$, and supports 
{\it fiducial Gaussian beams} with
beam radii at the mirrors $r_o = 4.70$ cm.  (This configuration differs from the current baseline
design for advanced LIGO; it is nearer an altered baseline that we recommend in Sec.\ 
\ref{sec:TEBL}.)
Our second configuration uses {\it fiducial MH mirrors}
and {\it fiducial mesa beams} with beam radii (defined in Sec.\ \ref{sec:MesaFields}) 
$D=4b = 10.41$ cm.  Our comparative conclusions for these two configurations are 
the following
(see
Sec.\ \ref{sec:Practical} for greater detail).  {\em [Note: if we had compared the baseline Gaussian-beam 
configuration with a mesa-beam configuration with the same diffraction losses, we would have reached
approximately the same conclusions; see, e.g., the paragraph following Eq.\ (\ref{alphanMH}).]}
\begin{enumerate}
\setcounter{enumi}{1}
\item
Among those parasitic optical modes of a perfect arm cavity
that are not strongly damped by diffraction losses, the parasite closest in frequency to the
desired TEM00 mode is separated from it by 0.099 of the free spectral range
in the 
fiducial 
Gaussian-beam case, and by 0.0404 of the free spectral range in the
fiducial
mesa-beam case; see Sec.\ \ref{sec:ParasiticModes}.  
This factor $\sim 2$ smaller mode spacing leads to a modestly
greater sensitivity of the mesa-beam interferometer to mirror tilt errors.

\item 
The interferometer's arm cavities are 
about
four 
times more sensitive to mirror
tilt when 
fiducial
MH mirrors are used than for  
fiducial
spherical mirrors.  
When all four cavity mirrors
are tilted through angles $\theta$ about uncorrelated
axes, the fractions of the carrier power driven into (dipolar) parasitic modes
inside the arm cavities, and driven out the dark port, are about $0.001 (\theta/0.01\;\mu{\rm rad})^2$
and $0.002 (\theta/0.01\; \mu{\rm rad})^2$, respectively
for 
our fiducial MH mirrors;
and $0.002 (\theta/0.035\; \mu{\rm rad})^2$ for 
our fiducial spherical mirrors;
Sec.\ \ref{sec:Tilt}.  This factor four
greater sensitivity is not a serious issue, since it turns out that the strictest constraints
on mirror tilt come from the signal recycling cavity (and, if a heterodyne output were
to be used, from the power recycling cavity), and not from the arm cavities; see 
items \ref{item7} and \ref{item8}
below.

\item
The
interferometer's arm cavities  are about 
equally sensitive to
transverse displacements of their end test-mass mirrors (ETM's), whether 
the mirror shapes
are fiducial MH, or are fiducial spherical.  
Specifically: 
for uncorrelated displacements
of the two ETMs  through distances $s$, the fractions of the carrier power 
driven into (dipolar) parasitic modes inside the arm cavities, and driven out the interferometer's
dark port, are about 
$100 (s/1.0\text{mm})^2$ ppm and $190(s/1.0\text{mm})^2$ ppm, respectively, for
fiducial MH mirrors; and
$100 (s/1.3\text{mm})^2$ ppm and $190(s/1.3\text{mm})^2$ ppm for fiducial spherical mirrors. 
For details, see Sec.\ \ref{sec:Displacement}.

\item
For MH mirror figure errors with peak-to-valley height variations $\Delta z$ in the innermost 10 cm
by radius:  after the control system has optimized the mirror tilts, the fractions of the carrier power driven into parasitic modes inside the arm cavities,
and driven out the dark port, are about  $0.0008 (\Delta z/6\text{ nm})^2$ and 
$0.0015 (\Delta z/6\text{ nm})^2$, respectively; Sec.\ \ref{sec:FigureErrors}. 
For spherical mirrors, a figure error $\Delta z$ about twice as large drives the same power into parasitic 
modes --- and the relevant figure error is confined to a smaller central region of the mirror (about 7.3
cm radius for baseline spherical mirrors and 8 cm for fiducial spherical mirrors, compared to 10 cm
for fiducial MH mirrors).  
The measured mirror figure errors in the initial LIGO interferometers are
of order 
$\Delta z = 3$ to 6 nm peak to valley (i.e.\ 1 to 2 nm rms),
which suggests that the MH arm cavities' required figure 
errors may be achievable.  

\item
The most serious constraints on mirror tilt and on mirror figure accuracy
come not from the arm cavities but rather from the signal recycling (SR) cavity.
The SR cavity, 
power recycling (PR) cavity, 
and RF sideband cavity 
operate
approximately
in the geometric optics regime and thus are nearly insensitive to
whether one uses MH or spherical mirrors; Sec.\
\ref{sec:RecyclingCavities}.  As a result, \emph{by switching from 
spherical to MH mirrors, one pays only a small penalty, in terms of 
mirror tilt constraints and figure-error constraints.}  
\item 
\label{item7}
More specifically, the most severe constraints
on tilt and figure error arise from the driving of signal power into 
parasitic modes when the signal light passes through the SR cavity.  To keep the
resulting 
loss of signal strength 
below one per cent in the standard wideband
advanced LIGO interferometers, it is necessary to constrain
the magnitude $\theta$ of the vectorial tilts of the input test-mass mirrors (ITM's) and signal recycling mirror
(SRM) to 
$\theta^{\rm sph}_{\rm WB} \lesssim 0.024 \; \mu$rad (for the fiducial spherical mirrors) and
$\theta^{\rm MH}_{\rm WB} \lesssim 0.016 \; \mu{\rm rad}$ (for the fiducial
MH mirrors).  
When an
advanced interferometer is narrowbanded at 
$f\simeq 500$ Hz or $\simeq 1000$ Hz, the constraint must be tighter:
$\theta^{\rm sph}_{\rm NB} \lesssim 0.011 \; \mu$rad, and
$\theta^{\rm MH}_{\rm NB}
\lesssim 0.007\; \mu{\rm rad}$.   
These are approximately the same as the experimentally
achieved \cite{fritschel} 
constraints on LIGO-I tilt arising 
from
the 
RF 
cavity in the absence of an output mode cleaner, 
and the same as the baseline constraints on tilt in advanced LIGO arising from
misalignment of the input laser beam\cite{muller}.  
If there were no
output mode cleaner in advanced LIGO and heterodyne readout were used
in place of the baseline homodyne readout, then 
the constraint on tilts in the 
RF 
cavity 
(due to mode mixing for the  RF sideband light used in the readout)
would be about the same as that for
narrowband
interferometers in the SR cavity.  For the 
baseline 
homodyne readout,
no such 
RF 
constraint arises.
The reduction in signal strength due to all the above (item \ref{item7}) tilts
scales as $\theta^2$; and we estimate that
the above (item \ref{item7}) 
constraints are inaccurate by a factor $\lesssim 2$  due to 
ignoring correlations in the overlaps of certain parasitic modes, and for the
narrowbanded interferometers, due to inaccuracy of the geometric optics approximation
in the SR cavity.  For details of all these issues, see Sec.\  \ref{sec:TiltShot}.
\item 
\label{item8}
We characterize the analogous 
constraints on mirror figure error by the \emph{peak-to-valley} fluctuations in the mirror height
in the central regions of the mirrors (regions enclosing 95 per cent of the light
power; radius $\simeq10$ cm for 
fiducial
MH mirrors and $\simeq 8$ cm for 
fiducial 
spherical mirrors), 
with the fluctuations averaged
over $\sim 3$ cm (an averaging produced by breakdown of geometric optics in
the SR cavity).  Our estimated constraints for one per cent 
loss of signal strength
are
$\Delta z_{\rm WB}  \lesssim 2.0$ nm for wideband advanced LIGO interferometers and $\Delta z_{\rm NB} \lesssim 1.0$ nm for narrowband, 
independently of whether the mirrors are MH or 
spherical---though the region over which the constraints must be applied is different, 10 cm radius for MH and 8 cm for 
spherical.
The 
loss of signal strength
scales as $\Delta z^2$, and our estimated 
constraints might be inaccurate by as much as a factor $\sim 3$ due to exploring only
one representative shape for the figure errors, due to overlaps of certain parasitic modes, and for the narrowbanded interferometer due to inaccuracy of the geometric optics approximation in the SR cavity.
These are approximately the same constraints as arise (in our calculations) from the 
RF
cavity in LIGO-I,  in the absence of an output mode cleaner.  If there were
no output mode cleaner in advanced LIGO and heterodyne readout were used, 
then
the constraint on 
figure errors 
in the 
RF 
cavity (due to mode mixing for the 
RF sideband light used in the readout) would be about the same as that 
for       
wideband interferometers in the SR cavity.
For details of these conclusions, see Sec.\ \ref{sec:FigureShot}.
\end{enumerate}

Among all the constraints on mirror errors that arise from our modeling, the most
serious are the last ones 
(items \ref{item7} and \ref{item8}): 
SR-cavity-induced 
constraints on mirror figure errors to avoid a one per cent 
loss of signal strength.  
These constraints are nearly independent of whether the mirrors
are spherical or MH.  These constraints would be relaxed if the SR cavity were
made less degenerate.  This could be achieved by  shaping the fronts of
the ITMs as lenses that bring the light (Gaussian or mesa) to a focus somewhere
near the SR mirror --- and also near the PR mirror.

Because MH mirrors and their mesa beams produce such a great (factor 3) reduction of
thermoelastic noise power, and they increase the sensitivity to mirror errors by
only modest amounts, they have been adopted as options for advanced LIGO, and they
may be of value for LIGO's future international-partner interferometers.  In
Sec.\ \ref{sec:Future} we describe some of the future research that is needed in
order to firm up our understanding of the pros and cons of MH mirrors and mesa beams.

\subsection{Notation}
\label{sec:Notation}

We here summarize some of the notation used in the remainder of this paper.  The numerical
values are for advanced LIGO interferometers, including sapphire test-mass substrates, with
the sapphire idealized as isotropic (its properties averaged 
over directions).\footnote{\label{fn:anistropy}
The influence of mechanical anisotropies and
of heat-conduction anisotropies on the thermoelastic noise have not yet been investigated, but
could be large enough to be important, since the relevant parameters vary with direction by amounts
of order $\pm 5\%$ for the Young's modulus $E$, $\pm 25\%$ for $1-2\sigma$, $\pm 5\%$
for the heat conductivity $\kappa$, and $\pm 5\%$ for the thermal expansion coefficient 
$\alpha_l$.
}

\begin{description}

\item[$b$:] Diffraction lengthscale $b=\sqrt{\lambda L/2\pi} = 2.603$ cm for
light 
with wavelength $\lambda=1.064\; \mu$m  
in the $L=4$km LIGO beam tubes; 
equal to a symmetric Gaussian beam's 
minimum possible radius at the end mirrors
(the radius at which the power flux has dropped to $1/e$ of its central value).
\item[$C_V$:] Specific heat of test-mass substrate per unit mass at
  constant volume 
[$7.9\times 10^6 
\text{erg g}^{-1} \, \text{ K}^{-1}$]

\item[$D$:] Mesa beam radius
[units cm]; 
Eq. (\ref{standard})

\item[$E$:] Young's modulus of test-mass substrate 
[$4\times10^{12}\;  
\text{dyne}\, \text{cm}^{-2}$]

\item[ETM:] End test mass of an arm cavity

\item[$H$:] Thickness of test mass
[units cm]

\item[$f$:] Gravitational-wave frequency at which noise is evaluated
[units Hz]

\item[$g$:] 
Arm cavity's g-factor, $g=1-L/{\mathcal R}_c$ where $L$ is arm length and
${\mathcal R}_c$ is the (identical) radius of curvature of its mirrors

\item[$\mathcal F$:] Finesse of an optical cavity

\item[$I$:] Noise integral for a test mass, Eq.\ (\ref{IA})
\item[ITM:] Input test mass of an arm cavity
\item[$k$:] Wave number, equal to $2\pi/\lambda$

\item[$k_B$:] Boltzmann's constant 
   [$1.38\times 10^{-16} \; 
\text{erg} \, \text{K}^{-1}$]

\item[$L$:] Interferometer arm length [$4 \times 10^5 \text{cm}$]

\item[${\cal L}$:] Diffraction loss in a single reflection off a mirror

\item[$M$:] Mass of test mass [$4\times 10^4 \text{g}$]

\item[{\bf PRM}:] Power recycling mirror

\item[$r$] Radius in transverse plane

\item[$r_o$:] 
Radius, on test-mass face, at which the {\em intensity} of a  Gaussian light beam has
dropped by a factor $1/e$ from its central value [$r_o = 4.23 \text{ cm} = 1.63 b$ for baseline design].  
Note: many LIGO papers, following Siegmann\cite{Siegmann71}, define the beam radius to be $\sqrt{2}$ 
larger, so it is the radius at which the beam's amplitude has dropped by $1/e$.
 
\item[$P_n$:] Fraction of interferometer's light power in mode $n$

\item[$R_p$:] The physical radius of a test mass [$15.7 \text{cm}$] 
\item[$R$:] The radius of the mirror coated onto a test mass [equal to $R_p$ or $R_p-8$mm]; also, the power reflectivity of a mirror

\item[${\mathcal R}_c$] 
Radius of curvature of a spherical mirror [units cm]

\item[$s$:] Transverse displacement of an arm cavity's ETM

\item[$S_h(f)$:] Spectral density of noise (thermoelastic or other) for detecting a gravitational wave $h$ with optimal direction and polarization
[units Hz$^{-1}$]

\item[{\bf SRM}:] Signal recycling mirror

\item[$T$:] Temperature of test-mass substrate [$300 \text{K}$]

\item[$u$:] Electric field of some light mode or superposition of modes (renormalized to unit norm, $\int |u|^2 d$Area$=1$); usually evaluated at the transverse plane tangent to an ITM mirror face, with the light propagating away from the ITM.  Subscripts identify the mode.
\item[$U$:] Unnormalized electric field of some light mode.
\item[$v$:] Same as $u$: Unit-normed electric field of some light mode or superposition of modes.

\item[$\Delta z$:] The peak-to-valley mirror deformation (mirror figure error) 
in a mirror's central region

\item[$\alpha_l$:] Substrate's coefficient of linear thermal expansion

  [$5.0 \times 10^{-6}$ K$^{-1}$]

\item [$\alpha_{1,2}$:] Amplitude of excitation of an arm cavity's parasitic mode $u_{1,2}$ by a tilt of the cavity's ETM; Eq.\ (\ref{u0prime})

\item [$\beta_1$:] Amplitude of excitation of the parasitic mode $v_1$ by mirror figure errors; Eq. (\ref{uDeformed})
\item[$\gamma_0$] Overlap of arm cavity's fundamental mode $u_0$ with Gaussian mode 
$u_d$ that drives it; Eq.\ (\ref{gamma0})
\item[$\delta_\ell$] Fraction of the light power of some perturbed field $u'_\ell$ that is in parasitic modes; Eq.\ (\ref{deltadef})
\item[$\kappa$:] Thermal conductivity of test-mass substrate 
[$3.3 \times 10^6$ erg s$^{-1}$ cm$^{-1}$ K$^{-1}$]
\item[$\lambda$:] Wavelength of laser light [$1.064 \; \mu{\rm m}$]; also, in Sec.\ \ref{sec:SignalInArm}, a function appearing in the analysis of the signal recycling cavity.

\item[$\rho$:] Density of test-mass substrate [$4 \text{g} \, \text{cm}^{-3}$]; also, amplitude reflectivity of signal recycling mirror

\item[$\sigma$:] Poisson ratio of test-mass substrate [$0.23$]

\item[$\theta$:] Angle of mirror tilt

\item[$\theta_{\rm diff}$:] 
Diffraction angle of a Gaussian beam (the Gaussian beam's opening angle if it were to continue onward through the mirror to large distance); $\theta_{\rm diff} = \lambda/\pi w_o = 2[(1-g)/(1+g)]^{1/4} b/L$, where $w_o$ is the mode's waist radius (radius at which
the {\em amplitude} has dropped by $1/e$ from its central value).

\item[$\Theta$:] expansion (fractional volume change) of substrate

\item[$\omega = 2\pi f$:] Angular frequency, corresponding to the frequency 
$f$ at which the noise $S_h$ is evaluated

\item[$\zeta_{1,2}$:] Amplitude of excitation of an arm cavity's parasitic mode by transverse displacement of the ETM; Eq. (\ref{uprimedisplace}) 

\end{description}

\section{Mexican-Hat Mirrors and The Mesa Modes They Support \label{sec:MH}}

In this paper we study a specific variant of a mesa light beam
and the MH mirrors that support it.  We believe this variant to be near optimal for
reduction of thermoelastic noise, but we have not carried out the (rather complex) analysis
required to prove optimality.

\subsection{Mesa Fields}
\label{sec:MesaFields}

The flat-topped (mesa-shaped) eigenmode of an interferometer arm cavity, which
we seek to construct, must have an intensity
distribution that is nearly flat across most of the light beam, and that then falls as 
rapidly as possible (constrained by diffraction effects) at the beam's edges.
Moreover, if (as in baseline advanced LIGO)
the cavity's input test mass (ITM) and end test mass
(ETM) have the same physical dimensions, then 
to minimize the thermoelastic noise at fixed net diffraction loss, 
the beam should be symmetric about
the arm cavity's mid plane, so its beam radii $D$ are the same on the two
mirrors.  Otherwise [since $S_h^\text{TE}  \propto 1/D^3$ approximately, and 
diffraction losses  increase exponentially  rapidly with increasing $D$; Eq.\ (\ref{Lclipmesa})], 
the mirror
with the reduced beam radius and smaller diffraction loss
will have its thermoelastic noise power increased, 
while that with the enlarged beam radius and larger diffraction loss will have its
noise power decreased more modestly, leading to a net noise increase.

The fastest possible falloff, for light in an optical cavity of length $L$, is
that on the edge of the {\em minimal Gaussian beam} --- the Gaussian beam whose radius
increases by a factor $\sqrt 2$ in going from the beam waist (at the cavity's center
plane) to the cavity's end mirrors.  This minimal Gaussian, at the mirror planes,
has the following (unnormalized) form
\begin{equation}
U_{ \hbox{min Gauss}}(r) 
= \exp \left [ \frac{-r^2(1+i)}{2 b^2}  \right ]\;,
\end{equation}  
where 
\begin{equation}
b=\sqrt{L/k} = \sqrt{\lambda L/2\pi} = 2.603 \hbox{ cm}\;,
\label{bdef}
\end{equation}
with $L = 4$ km the cavity length, $k = 2\pi/\lambda$ the wave number, and
$\lambda = 1.064 \mu$m the wavelength of the light.

A near-optimal flat-topped eigenmode, with near-minimal thermoelastic noise, will have
a constant intensity in the central region, and will fall off at its edge at approximately the 
same rate as
the edge of this minimal Gaussian.  To produce such an eigenmode, 
we superpose minimal-Gaussian fields, with a field density that is constant out to
a radius $r=D$ and then stops abruptly.  More specifically, our chosen
unnormalized eigenmode has the following form: 
\begin{eqnarray}
&&U(D,r) = \nonumber\\
&& \quad \int_{\mathcal C_D}  \exp\left[ - [(x-x_o)^2 + (y-y_o)^2][1+i]\over 2b^2\right] dx_o dy_o
\;,\nonumber\\
\label{standard}
\end{eqnarray}
where $r\equiv \sqrt{x^2+y^2}$ and
the integration is over a circle $\mathcal C_D$ of radius $D$:  $\sqrt{x_o^2+y_o^2} < D$. 

By carrying out the
$y_o$ integral in Cartesian coordinates, with $y=0$ and $x=r$, we obtain the
following expression for $U(D,r)$, which we have used in much of our numerical work: 
\begin{eqnarray}
U(D,r) & = & b \sqrt{-2\pi\over 1+i} \int_{-D}^{+D} dx_o \exp\left[ {-(x_o-r)^2(1+i)\over 2b^2} \right] \nonumber \\
&& \times \text{erfi}\left[ {\sqrt{D^2-x_o^2}\over b}\sqrt{-(1+i)\over 2}\right]\;.
\label{standard1}
\end{eqnarray}
Here $\text{erfi}(z) = \text{erf}(iz)/i$ is the imaginary error function.

By converting to circular polar coordinates and performing the angular integral, we obtain
the following simpler expression for our unnormalized eigenmode
\begin{eqnarray}
U(D,r) &=& 2\pi\int_0^D \exp\left[{-(r^2+r_o^2)(1+i)\over2b^2}\right]\nonumber\\
&&\times I_0\left[ {r r_o(1+i)\over b^2}\right]
r_o dr_o\;.
\label{new}
\end{eqnarray}
Here $I_0$ is the modified Bessel function of order zero.  Modes with other weightings
of the minimal-radius Gaussians can be obtained by inserting a weighting function $f(r_o)$
into the integrands of Eqs.\ (\ref{standard}) and (\ref{new}).  

In the Appendix 
we give some approximate formulae for $U(D,r)$  valid
at large radii.  These are useful for quick, clipping-approximation computations of diffraction losses.

The squared norm of $U(D,r)$ (the
area integral of its squared modulus) is given by the following
approximate formula, which is accurate to within a fraction of a 
per cent for $3.0 \lesssim D/b \lesssim 6.0$ (the regime of interest to us): 
\begin{equation}
N^2(D) \equiv \int_0^\infty |U(D,r)|^2 2 \pi r dr = 4.66 - 50.58D + 62.10D^2\;.
\label{Norm}
\end{equation}  
We denote by $u$ the normalized field on the mirror faces, and to distinguish
it from a Gaussian field, we sometimes will use a subscript ``mesa'':
\begin{equation}
u_\text{mesa}(D,r) = u(D,r) = {U(D,r)\over N(D)}\;.
\label{uMH}
\end{equation}

\subsection{Gaussian fields}

The advanced LIGO baseline design uses arm cavities with spherical mirrors, 
which have Gaussian modes whose field at the mirror plane is (cf. \cite{Siegmann71}) 
\begin{equation}
u_G(r,r_o) = {1\over \sqrt{\pi r_o^2}} \exp\left[ -{r^2\over 2 r_o^2} \left(1-i{
b^2\over r_o^2 + \sqrt{r_o^4 - b^4}}\right) \right]\;.
\label{uGdef}
\end{equation}
Here $r_o$ is the beam radius (at which the energy flux falls to $1/e$ of its
central value).  From the phase of this field one can read off the radius
of curvature of the mirrors
\begin{equation}
{\cal R}_c = L \left({r_o\over b}\right)^2 \left[ \left({r_o\over b}\right)^2
+ \sqrt{\left({r_o\over b}\right)^4 - 1}\right]\;,
\label{Rc}
\end{equation}
and thence the arm cavity's g-factor
\begin{equation}
g=1-L/{\mathcal R}_c\;.
\nonumber
\end{equation}

\subsection{Diffraction Losses}

In the baseline design of an advanced LIGO interferometer \cite{ALDesign}, 
the test masses are cylinders
whose faces are coated with dielectric mirrors out to a radius $R$
that is 0.8cm less than the cylinders' physical radii 
\begin{equation}
R = R_p - 0.8 \text{cm}\;.
\label{RRp}
\end{equation}
We shall explore MH mirrors that are coated in this same manner, $R=R_p - 0.8$ cm and also
MH mirrors that are coated all the way out to the test-mass edges, $R=R_p$.
The diffraction losses in each reflection of a cavity mode off a mirror are
given, approximately, by the {\it clipping approximation}
\begin{equation}
{\cal L}_\text{clip} = \int_r^{\infty} |u(r)|^2 2\pi r dr\;.
\label{calLclip}
\end{equation}
Here $u(r)$ is the normalized field [$u_\text{mesa}(D,r)$ for a mesa mode
and $u_G(r_o,r)$ for a Gaussian mode with infinite mirrors, $R=\infty$].

In actuality, the mirrors' edges at $r=R$ modify the field thereby
causing the true diffraction losses to differ from this clipping formula.
The true diffraction losses have been computed by OSV \cite{OSV} via a numerical solution of
the eigenequation for the cavity modes, and independently by D'Ambrosio \cite{dambrosio} 
using an FFT code to propagate light in the cavity.  The results are
\begin{eqnarray}
{\cal L}_0 &\simeq& 0.85 {\cal L}_\text{clip} \hbox{ for mesa modes}\;, \nonumber\\
{\cal L}_0 &\simeq& 2.5 {\cal L}_\text{clip} \hbox{ for Gaussian modes}\;,\;
\label{calL0}
\end{eqnarray}
in the parameter regime of interest --- though the numerical coefficients
0.85 and 2.5 can oscillate substantially as the beam radii and mirror radii
are changed.  When we need high-accuracy diffraction losses (e.g.\ in portions
of Sec.\ \ref{sec:Practical}), we compute them with care using the cavity
eigenequation \cite{OSV}) or FFT code \cite{dambrosio}). 

\subsection{Mirrors and Normalized Flux for Mesa and Gaussian Modes}

\begin{figure}
 \includegraphics[width=3.4in]{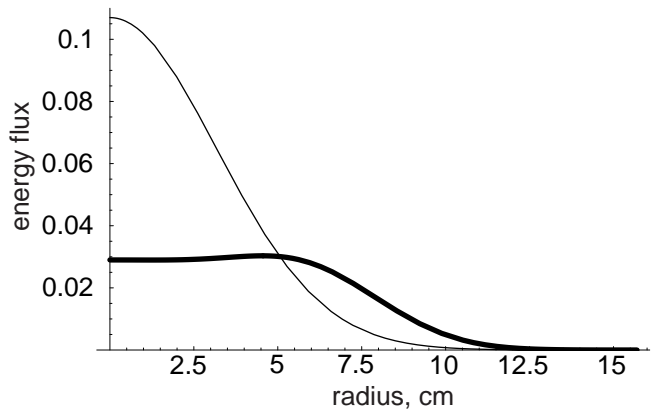}%
 \caption{The power distributions of: (i) the Gaussian mode for the baseline mirrors
with coated mirror radius $R=15.7$ cm and beam radius
$r_o = 1.73b=4.50$ cm (thin curve), which has diffraction loss per bounce 
${\cal L}_0 = 10$ ppm; and
(ii) the mesa mode with $D = 3.73b=9.71$ cm (thick curve) which, for this same coated 
mirror radius
$R=15.7$ cm, has the same diffraction loss per bounce ${\cal L}_0 = 10$ ppm.
\label{fig:MHGaussModes}}
 \end{figure}

\begin{figure}
 \includegraphics[width=3.4in]{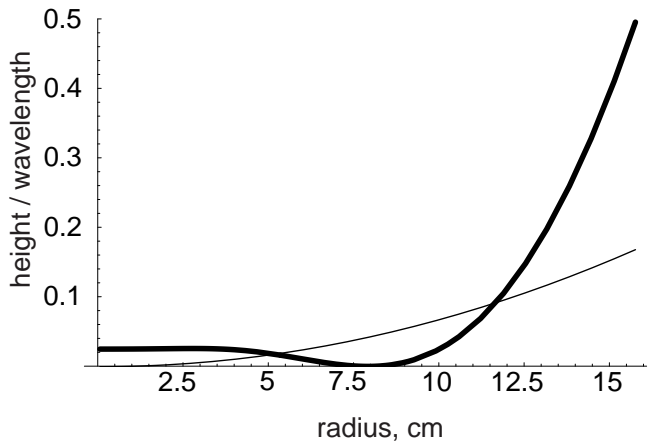}%
 \caption{The shape of the spherical mirrors (thin curve) and MH mirrors (thick curve)
that support
the arm-cavity Gaussian mode (thin curve) and mesa
mode (thick curve) of Fig.~\ref{fig:MHGaussModes}.  The height is measured in units
of the wavelength of the light, $\lambda = 1.064 \mu$m.
\label{fig:MHGaussMirrors}}
 \end{figure}

The baseline design for advanced LIGO interferometers
has mirror radii $R=14.9$ cm and Gaussian beam-spot radii $r_o = 1.63b = 4.23$ cm,
corresponding to a diffraction loss of ${\cal L}_0 = 
10$ ppm
and a mirror
radius of curvature ${\cal R}_c = 53.7$ km.  The normalized energy flux
$|u_G(r_o,r)|^2$ for this baseline Gaussian field is shown in Fig.\
\ref{fig:MHGaussModes}, 
and the shape of the mirror (segment of a \emph{sphere} with radius 53.7 km)
is shown in Fig.\ \ref{fig:MHGaussMirrors}.   

A cavity made from MH mirrors with the baseline radius $R=14.9$ cm and the
baseline diffraction losses ${\cal L}_0 = 10$ ppm has a mesa beam radius
$D = 3.43 b = 8.92$ cm [computed from Eqs.\ (\ref{calL0}) and (\ref{calLclip})].  
The normalized
energy flux $|u_\text{mesa}(D,r)|^2$ for this mesa field 
is shown in Fig.\ \ref{fig:MHGaussModes}.  
Notice how flat the top of this intensity profile is, and how much like
a mesa the profile is shaped, 
and notice the contrast with the Gaussian profile.    

The surfaces of the MH mirrors coincide with the mesa field's surfaces of
constant phase; i.e., their height $\delta z$ as a function of radius $r$
is given by 
\begin{equation}
k \delta z = \text{Arg}[ u_\text{mesa}(D,r)]\;,
\label{delta z}
\end{equation}
where $k = 2\pi/\lambda$ is the light's wave number.  This MH mirror shape
is shown in Fig.\ \ref{fig:MHGaussMirrors}.  
Notice the shallow bump in the middle and the
flaring outer edges.  This bump and flare resemble a Mexican hat (sombrero) and
give the MH mirror its name. 

\section{Thermoelastic Noise and Neutron-Star Binary Range for Mesa-Beam 
Interferometers }
\label{sec:TENoise}

\subsection{Thermoelastic Noise}

\subsubsection{Quantifying the thermoelastic noise: the noise integral and the NS/NS range}

Building on the seminal work of Braginsky, Gorodetsky and Vyatchanin \cite{BGV}, Liu
and Thorne \cite{LiuThorne} have used Levin's \cite{Levin} direct method to derive
the following 
formula for an interferometers' thermoelastic noise in terms of a noise integral $I_A$:
\begin{equation}
S_h(f) = 16 \kappa k_B \left( {\alpha_l E T \over C_V (1-2\sigma) \rho \omega L }\right)^2 \bar I\;;
\label{Sh}
\end{equation}
where $\bar I$ is the average, over the four test masses, of the thermoelastic noise integral,
\begin{equation}
\bar I = {1\over 4} \sum_{A=1}^4 I_A\;, \quad
I_A = {1\over F_o^2} \int_{V_A} (\vec\nabla \Theta)^2 d\text{volume}\;;
\label{IA}
\end{equation}
cf.\ Eqs.\ (3), (4) and (13) of \cite{LiuThorne}.
In Eq.\ (\ref{Sh}),
the notation is as spelled out in Sec.\ \ref{sec:Notation}, and we use
numerical values (shown in Sec.\ \ref{sec:Notation})  
that assume the test-mass substrate is sapphire, idealized
as an isotropic material.$^{
\ref{fn:anistropy}
} $
In Eq.\ (\ref{IA}), $\Theta$ is the expansion (fractional volume change) 
inside the test-mass substrate, produced
by a static force with magnitude $F_o$ and with profile identical to that of the light beam's 
intensity 
distribution over the test-mass face (e.g., Fig.\ \ref{fig:MHGaussModes}), and the integral is over
the volume $V_A$ of test-mass $A$.  Note that the dimensions of $I_A$ and thence of $\bar I$ are 
length/force$^2 = \text{s}^4 \text{g}^{-2} \text{cm}^{-1}$.

Equation (\ref{Sh}) shows that the frequency dependence of the thermoelastic noise is independent of the mirror
shape and test-mass shape; it always has the same slope as the SQL (except in testbed
systems with tiny mirrors and light beams \cite{LiuThorne}, which are
irrelevant in this paper).  As a result, the thermoelastic noise produced by an advanced LIGO
interferometer whose mirrors
have some chosen shapes, divided by the thermoelastic noise of the baseline advanced LIGO interferometer,
is equal to the ratio of the two interferometers' 
noise integrals
\begin{equation}
[S_h(f)/
S_h^\text{BL}(f)]_\text{TE} = \bar I/\bar I_\text{BL}
\label{ShRatio}
\end{equation}
This motivates our use of $\bar I/\bar I_\text{BL}$ as one measure of a candidate interferometer's
thermoelastic noise.

We shall also use a second measure: 
The distance to which the LIGO network can detect NS/NS binaries, with
network amplitude signal-to-noise ratio 8; we call this the network's {\em range} for NS/NS
inspirals.  This three-interferometer
network range
is larger by a factor $\sqrt{3} = 1.732$
than the single-4km-interferometer NS/NS range that is often used by the LIGO community and
that
is encoded into the ``BENCH'' LIGO software \cite{bench}.  In computing the network range for advanced LIGO with Gaussian or mesa beams, 
we assume that: (i)
the three advanced LIGO interferometers (all with $L= 4$ km) are all being operated
with signal-recycling-mirror parameters that maximize the range
(the operation mode 
tentatively planned for the first year of advanced LIGO observations), (ii) they all   
incorporate identical sapphire mirrors with the same 
shapes and beam sizes, (iii) all thermal noises are negligible except thermoelastic noise,
and (iv) the remaining interferometer parameters have their baseline advanced LIGO values \cite{ALDesign}
(e.g., the circulating power in each arm is 830 kW).     {\em For the baseline advanced LIGO
design with Gaussian beams, 
the single-4km-interferometer NS/NS range, computed under these assumptions, is 200 Mpc, and the network NS/NS
range is $200 \sqrt{3} = 346$ Mpc.} 

Since the only noise source we change, in going from one candidate interferometer design to another, 
is the thermoelastic noise, the NS/NS range must be some function of $[S_h(f)/
S_h^\text{BL}(f)]_\text{TE} = \bar I/\bar I_\text{BL}$.  

\begin{figure}
 \includegraphics[width=3.5in]{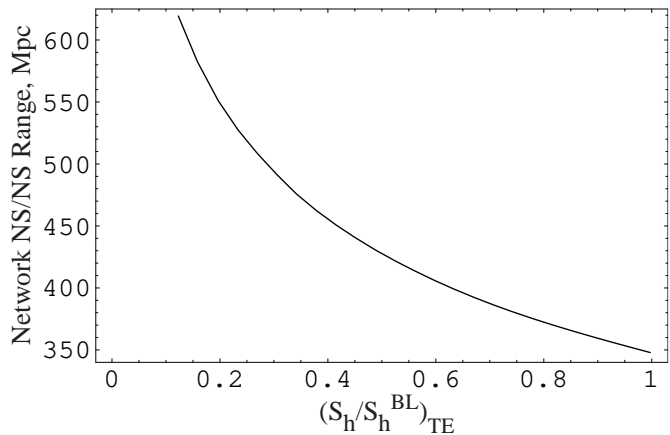}%
 \caption{
The LIGO network NS/NS range as a function of the thermoelastic noise power,
in units of the baseline thermoelastic noise, $(S_h/S_h^\text{BL})_\text{TE}$.
For each thermoelastic noise level, the advanced LIGO interferometer's optical
parameters (homodyne readout phase and signal-recycling mirror)
are optimized to produce the greatest possible NS/NS range.  The optimization
has been performed for us by A.\ Buonanno and Y.\ Chen (private communication),
assuming that the only significant noise sources are thermoelastic noise and
optical (unified quantum) noise.
\label{fig:Range}}
 \end{figure}

Buonanno and Chen (private communication) have performed the optimization of the 
advanced LIGO optical-noise parameters (the homodyne detection phase and the position and 
reflectivity of the
signal recycling mirror), as a function of the thermoelastic-noise level, to produce for us
a curve of optimized NS/NS signal-to-noise ratio $S/N$ as a function of 
$(S_h/S_h^\text{BL})_\text{TE}$.  From that $S/N[ (S_h/S_h^\text{BL})_\text{TE}]$, we have
computed the corresponding network range, $ (346\hbox{ Mpc}) \times  (S/N) (S/N)_\text{BL}^{-1}$ as a
function of thermoelastic noise. We show that range in Fig.\ \ref{fig:Range}.

A third measure of a candidate interferometer's performance is the ratio of its network event rate for
NS/NS binaries to that of the baseline advanced LIGO network.  Since the NS/NS binaries are very
extragalactic, their event rate scales as the range cubed,
\begin{equation}
\hbox{Rate/Rate}_\text{BL} = (\hbox{Range/346 Mpc})^3\;.
\label{RateRatio}
\end{equation}

In a companion paper \cite{OSV}, OSV 
evaluate the thermoelastic noise integral $I_A$ numerically for a variety of test-mass
shapes and beam radii.    Here we summarize the most important conclusions of those
computations and their implications for our three performance parameters: 
$[S_h(f)/ S_h^\text{BL}(f)]_\text{TE}$ [Eq.\ (\ref{ShRatio})], 
NS/NS range [Fig.\ \ref{fig:Range}], and NS/NS event rate [Eq.\ (\ref{RateRatio})].

\subsubsection{Baseline test masses with spherical mirrors and Gaussian beams}
\label{sec:TEBL}

The baseline design of an advanced LIGO interferometer entails four identical test masses:
sapphire cylinders with physical radii $R_p=15.7$ cm, coated-mirror radii
$R = R_p - 0.8$ cm, thickness $H = 13$ cm,
density $\rho = 4$ g  $\text{cm}^{-3}$ and mass $M = 40$ kg; and the baseline
light beam at the test-mass face is Gaussian with beam radius $r_o = 4.23$ cm  $= 1.63 b$
so the diffraction loss per bounce is ${\cal L}_0 = 10$ ppm.
For this baseline beam and test mass, OSV \cite{OSV} find for the value of the noise integral 
\begin{equation}
I_\text{BL} = 2.57 \times
10^{-28} \text{s}^4 \text{g}^{-2} \text{cm}^{-1}\;.
\label{IBL}
\end{equation}

We advocate extending the mirror coating out to the test-mass edge so $R=R_p = 15.7$ cm, and increasing the Gaussian
beam radius correspondingly, to 
$r_o = 4.49 \text{cm} = 1.725b$, 
so the diffraction losses are still
10 ppm.  With this beam expansion, OSV find 
(Table I of \cite{OSV})
that the thermoelastic noise is 
reduced to $ S_h/S_h^\text{BL} = 
I/I_\text{BL} = 0.856$ \cite{OSV}, from which we deduce (via Fig.\ \ref{fig:Range})
that the range for NS/NS binaries is increased from 346 Mpc to 364 Mpc, and the NS/NS event
rate is increased by a factor $(364/346)^3 = 1.16$; see Table \ref{tbl:NSRange}.

\subsubsection{Cylindrical test masses with MH mirrors and mesa beams}

OSV have computed the thermoelastic noise integral $I$ for cylindrical test masses with mesa
beams.  The test masses' 
volumes were held fixed at the baseline value of $10^4$ cm$^3$ (masses fixed at 40 kg), 
while their physical radii $R_p$
and thicknesses $H$ were varied.  For each choice of $R_p$, two coated-mirror radii were chosen,
$R=R_p - 8$ mm (the baseline choice) and $R=R_p$ (our proposed expansion of the coating). 
In all cases the mesa beam radius $D$ was that value for which the diffraction losses are 10 ppm
per bounce inside the cavity.  

To within the accuracy of their computations, $\sim 0.5$ per cent,  OSV 
(Table II of \cite{OSV}) 
found that 
the thermoelastic noise integral $I$ is minimized when the test-mass dimensions have their baseline
values, $R=15.7$ cm, $H=13$ cm.  In other words, the optimal test-mass shape is the same for mesa
beams as for Gaussian beams.  The optimized (10 ppm diffraction loss) radii $D$ for the mesa
beams, and the values of our three measures of interferometer performance are shown in
Table \ref{tbl:NSRange}, in two cases: for mirrors coated out to $R=R_p - 8$ mm (the baseline
choice), and coated out to $R=R_p$.

\begingroup
\squeezetable
\begin{table}
\caption{\label{tbl:NSRange} Optimized light-beam configurations, their
thermoelastic noise compared to the baseline
(from Table IV of \cite{OSV}), 
their neutron-star binary range, and their
event rate for NS/NS inspiral divided by the baseline rate.  All test masses are assumed
to  be cylinders with the baseline advanced LIGO dimensions: physical 
radius $R_p=15.7$ cm and thickness $H=13.0$ cm.  The beam radii $r_o$ and $D$ are chosen 
so that the diffraction loss per bounce in the arm cavities is 10 ppm.
}
\begin{ruledtabular}
\begin{tabular}{lldcr}
Coated Radius & Beam Shape & 
\multicolumn{1}{c}{$\left(\frac{S_h}{S_h^\text{BL}}\right)_\text{TE}$} 
   & NS/NS & $\frac{\text{Rate}}{\text{Rate}_\text{BL}}$ \\
 & and Radius &  & Range  & \\
\hline
$R=R_p-8\text{mm}$ & BL: Gaussian & & \\
 &  $r_o = 4.23\text{cm}$ & 1.000 & 346 Mpc & 1.00 \\
$R=R_p-8\text{mm}$  & mesa & &\\
 &  $D/b = 3.43$ & 0.364 & 465 Mpc & 2.42 \\
\\
$R=R_p$ & Gaussian & &\\
 &  $r_o = 4.49\text{cm}$ & 0.856 & 364 Mpc & 1.16 \\
$R=R_p$ & mesa  & &\\
 &  $D/b = 3.73$ & 0.290 & 497 Mpc & 2.97 \\
\end{tabular}
\end{ruledtabular}
\end{table}
\endgroup

As is shown in the table, \emph{switching from Gaussian beams to mesa beams
reduces the thermoelastic noise $S_h \propto \bar I$ by about a factor 3; it increases the
NS/NS range from 346 Mpc to 465 Mpc if $R=R_p-8$ mm, and 497 Mpc if $R=R_p$; and it increases
the NS/NS event rate by a factor $(465/346)^3 = 2.42$ if $R=R_p-8$ mm, and 
to $(497/364)^3 = 2.55$ if $R=R_p$.}  

\subsubsection{Conical test masses}

By switching from cylindrical test masses to frustums of cones, with the same test-mass volume, 
one could further reduce, substantially, the thermoelastic noise and  increase the
NS/NS range and rate.  For detailed explorations of this, see 
Tables III and IV and Appendix I of OSV \cite{OSV}.

We do not discuss this possibility in the present paper because the current technology for growing
sapphire boules, from which to cut the advanced LIGO test masses, places a tight limit on the 
test-mass physical radius $R_p$.  It cannot be much larger than the baseline $R_p = 15.7$ cm; and for
that maximum radius, and test-mass volumes of order the baseline $10^4$ cm$^3$, the optimal test-mass shape
is cylindrical, with the baseline dimensions \cite{OSV}.   

When it becomes possible, in the future, to grow larger sapphire boules, it might be worth
considering test masses with 
frustum 
shapes \cite{OSV}.
 
\section{Sensitivity to Mirror Tilts, Displacements and Figure Errors \label{sec:Practical}}

The MH mirror figure (Fig.\ \ref{fig:MHGaussMirrors}) is somewhat flatter than the baseline spherical mirror in its
central 10 cm of radius where 95 per cent of the light power resides,
but much more curved in its outer $\sim 6$ cm.  One might worry that the central flatness
will cause a mesa-beam interferometer to be unacceptably sensitive to mirror-tilt-induced, 
mirror-displacement-induced and figure-error-induced
mixing of parasitic modes into the light beam's fundamental, mesa mode.  We have 
investigated this mode mixing and find that it is a modest problem, not a severe one.  We 
describe our investigations and conclusions in this section.  They have been described
previously in our internal LIGO report \cite{DOSTVshortdoc}, and a short summary of
results was given
in Sec.\ \ref{sec:SummaryResults} above.

\subsection{Foundations for Investigation}
\label{sec:Foundations}

\subsubsection{Our tools of analysis}

Our analysis of mode mixing and its consequences is based on three independent sets of tools.
The first two sets are designed for studying the effects of mirror errors on the interferometer's
high-finesse arm cavities.  The third set is for analyzing the highly degenerate power-recycling
and signal-recycling cavities.

\emph{Our first tool set} (developed by Richard O'Shaughnessy with confirming calculations by Sergey Strigin and Sergey Vyatchanin,
and described in detail in OSV \cite{OSV})
is an \emph{integral eigenequation} for the modes of an arm cavity.  In the limit of infinite
mirror radii (i.e., neglecting diffraction losses), the cavity's
eigenmodes are orthonormal
when integrated over the transverse plane; this is true for MH mirrors, just as for
spherical mirrors 
(Appendix D of \cite{OSV}).
OSV have used their integral eigenequation
to compute the modes with untilted, undisplaced and undeformed mirrors and with both infinite and finite
radii.  O'Shaughnessy has then tilted, displaced and deformed the ETM of one arm
cavity and applied \emph{first- and second-order perturbation theory} to its eigenequation
to determine the tilt-induced and deformation-induced mode-mixing, the resulting fundamental eigenmode
of the cavity with tilted and deformed ETM, its response to the driving beam, and the tilt-induced
and deformation-induced power going out the interferometer's dark port.  The details of these
calculations are given in OSV \cite{OSV}.  In the following subsections we describe the
main results, we compare with computations via our second tool set, and we discuss the implications
for the use of MH mirrors in advanced LIGO interferometers.

\emph{Our second tool set} (developed by D'Ambrosio and described in her companion
paper \cite{dambrosio})  
is an adaptation of a standard  VIRGO and LIGO \emph{FFT code for
simulating interferometers}\footnote{See 
http://www.phys.ufl.edu/LIGO/LIGO/STAIC.html where it is referred to as the FFT Full Field Relaxation Code.
}
(a code originally designed by Patrice Hello and Jean-Yves Vinet, then
further develped by Brett Bochner and others; see \cite{bochner} and
references therein).
D'Ambrosio has used her adaptation of this code
to study the same arm-cavity phenomena that OSV have studied via the cavity eigenequation and
perturbation theory.  She presents the details of her computations and some associated 
perturbation theory analyses in Ref.\ \cite{dambrosio}.  In the following subsections we describe her
main results, we compare them with the OSV eigenequation results, and we discuss their
implications. 

\emph{Our third tool set} (developed by Thorne and described in Secs.\ 
\ref{sec:TiltShot} and \ref{sec:FigureShot} below) is designed
to deal with the influence of mirror errors on the interferometer's
power-recycling and signal-recycling cavities.  Because these cavities are highly degenerate, the
light trapped in them and traversing them is moderately well described by the tools of geometric
optics, augmented by approximate estimates of the impact of diffraction; so these are the foundation
for our third tool set.

\subsubsection{Fiducial mesa and Gaussian configurations}
\label{sec:fiducialconfigs}

When analyzing practical issues, we have carried out most of our computations for
{\it fiducial MH mirrors} that have slightly larger coated radii, 
$R_M=16$cm, than the baseline spherical mirrors (14.9 cm at present and 
15.7 cm if our recommendation to coat the outer 8 mm is followed), and we 
have used 
corresponding {\it fiducial mesa beams} 
with modestly larger diffraction
losses, $\mathcal L_0 = 18$ ppm, than the current baseline of about 10 ppm.
In evaluating practical issues in this
paper and its companions \cite{dambrosio,OSV}
we will compare with 
{\it fiducial spherical mirrors} and {\it fiducial Gaussian beams}
that have this same
enlarged coated radius $R_M = 16$ cm and diffraction losses $\mathcal L_0 =
18$ ppm.

Our fiducial mirror and beam parameters, then, are as follows:  
\begin{itemize}
\item
For the 
{\it fiducial spherical mirrors}: 
mirror radius $R_M=16$ cm (vs 15.7 cm for the current
baseline) and
Gaussian beam radius $r_o=4.70$ cm (vs.\ 4.23 cm for the current baseline),
corresponding to a mirror radius of curvature ${\mathcal R}_c = 83$ km (vs.\ 
53.7 km for the current baseline) and an arm-cavity g-factor 
$g=1-L/{\mathcal R}_c = 0.952$ (vs.\ 0.926 for the current baseline).
\item
For the {\it MH mirrors}: mirror radius $R_M=16$ cm and 
mesa beam radius parameter $D=
4b = 10.4$ cm, where $b= \sqrt{\lambda L/2\pi} = 2.60$ cm, with $\lambda=1.064$  $\mu$m
and $L=4$ km the light wavelength and arm length.   
\end{itemize}

[Note: if we had compared the baseline Gaussian-beam 
configuration with a mesa-beam configuration with the same diffraction losses, we would have reached
approximately the same conclusions as we obtain from comparing these fiducial configurations; 
see, e.g., the paragraph following Eq.\ (\ref{alphanMH}).]

\subsubsection{Reflection and transmission coefficients}
\label{sec:rtIrtE}

For both configurations, fiducial MH and 
fiducial
spherical, 
we have assumed for the ITM mirror
the advanced LIGO baseline power transmissivity $t_I^2 = 0.995$;
we have chosen an idealized, perfectly reflecting ETM; and for
simplicity we have assumed that the only losses are those due to diffraction, which 
we treat as a separate mathematical entity from the reflection and transmission      
coefficients.  Thus,  our power reflection and transmission for the ITMs and ETMs are 
\begin{eqnarray}
r_\text{I}^2 &=& 0.995\;, \quad t_\text{I}^2 = 0.005\;;
\nonumber \\
r_\text{E}^2 &=& 1.0\;, \quad t_\text{E}^2 = 0.
\label{rtF}
\end{eqnarray}
We have also assumed the baseline advanced LIGO transmissivities 
and reflectivities for the power-recycling (PR) and signal-recycling (SR) mirrors;  
see Secs.\ \ref{sec:TiltShot} and \ref{sec:FigureShot} below.

\subsection{Driving a Mesa-Beam Interferometer with a Gaussian Beam}
\label{sec:Drive}

One way to produce the desired mesa beams in the arm cavities is to drive the interferometer
with Gaussian-beam
light and let the arm cavities or a mode-cleaning cavity convert the light into mesa form.
One of us, ED'A
(Secs.\ II$\,$A and II$\,$B of \cite{dambrosio}), 
has identified the Gaussian beam that has the greatest overlap
with the mesa beam of an MH-mirrored arm cavity.  If one were to 
drive the MH arm cavity directly
with a Gaussian beam, this would be the driving beam one would want to use.  It
has a beam radius 
$r_{od} = 6.92$ cm,
compared to our 
fiducial
cavity's
beam radius at the ITM of $r_o = 4.70$ cm.
This Gaussian driving beam
$u_d(r)$ 
has an overlap
\begin{equation}
\gamma_0^2 \equiv |\langle u_0 | u_d \rangle|^2 \equiv \left|\int u_0^*u_d d\text{Area}\right|^2 =0.940
\label{gamma0}
\end{equation}
with the cavity's fundamental mesa mode $u_0$, which means that 94.0 per cent of the 
Gaussian driving-beam
light will enter the MH-mirrored cavity, and 6.0 per cent will get rejected.  
See Secs.\ II$\,$A and II$\,$B of \cite{dambrosio}, where $\gamma_0^2$ is
denoted $\mathcal C^2$.

\subsection{Parasitic Modes in Arm Cavities}
\label{sec:ParasiticModes}

It is useful to think of the MH mirrors as having two regions: a central region
with radius $\simeq 10$ cm, and an outer region 
extending from radius $\simeq 10$ cm to radius 16 cm.
In its central region, the 
fiducial 
MH mirror is much flatter than
the 
fiducial
spherical mirror; in its outer region, it is much more sharply curved;
see  Fig.\ \ref{fig:MHGaussMirrors}.
The flatness of the central region has led to concerns about         
degeneracies of modes and sensitivity to mirror tilts, displacements and figure errors.      

O'Shaughnessy, Strigin and Vyatchanin have all independently solved the integral
eigenequation for the modes of a LIGO arm cavity
with 
fiducial
MH mirrors.  They have found 
(Sec.\ VII$\,$A and Table V of \cite{OSV})
that \emph {\it among 
modes that are not strongly damped by diffraction
losses, the one closest in frequency to the fundamental 
TEM00 mode $u_0$ is the lowest
TEM01 mode (denoted $u_1$ below).  Its frequency separation from the 
fundamental 
is 0.0404 of the cavity's 
free spectral range, 
which is 
2.5 times smaller than for the 
fiducial
spherical-mirrored
cavity, $0.099 \times ($free spectral range).}  
Evidently, 
the sharp curvature
of the MH mirrors' outer region compensates sufficiently
for the flatness of their central region,
to prevent the parasitic modes' frequencies from becoming near-degenerate
with the fundamental.

\subsection{Mirror Tilt in Arm Cavities}
\label{sec:Tilt}

Our modeling  predicts that
\emph{mode mixing in the
arm cavities of a mesa-beam interferometer produced by tilt of the ETM's or ITM's should be 
of no serious consequence, if the tilt angles are kept below about $10^{-8}$ rad.}  In
the following subsections we summarize the calculations that lead to this conclusion.

\subsubsection{Parasitic mode mixing in arm cavities}

Two of us have computed the influence of a tilt of the ETM on the fundamental
mode of an arm cavity: ED'A has done
this using 
her 
FFT code, and RO'S has done it by applying perturbation
theory to the arm cavity's integral eigenequation.  The two computations agree
on the following predictions:

When the ETM is tilted through an angle $\theta$, the cavity's fundamental
mode gets changed from $u_0(r)$ to
\begin{equation}
u'_0(\vec r) = (1-\alpha_1^2/2) u_0(r) + \alpha_1 u_1(\vec r) + \alpha_2 u_2(\vec r)\;.
\label{u0prime}
\end{equation}
Here $\vec r$ is position in the transverse plane, $u_n$ are unit-norm 
superpositions of
modes of the 
perfectly aligned cavity ($\int |u_n|^2 d\text{Area} = 1$), $\alpha_n$ are mode-mixing coefficients that scale
as $\theta^n$, and our computations have been carried out only up through quadratic
order.  The maximum tilt that can be allowed is of order $10^{-8}$ radian, so 
we shall express our predictions for the $\alpha_n$ in units of $\theta_8 \equiv
\theta/10^{-8}$.

For our 
fiducial
spherical-mirrored cavities, $u_0$ is the (0,0) Hermite-Gaussian mode, 
$u_1$ is the (0,1) Hermite-Gaussian mode, $u_2$ is 
the (0,2) mode, and the dominant mixing coefficient $\alpha_1$ is
\begin{eqnarray}
\alpha_1^{\rm sph} &=&\frac{1}{\sqrt{2}(1-g^2)^{3/4}} 
  \left(\frac{\theta^{\rm sph}}{b/L}\right)  
  \label{alpha1BL}
  \\
 & =& \frac{\sqrt{2}}{(1+g)(1-g)^{1/2}} \left( \frac{\theta}{\theta_{\rm diff}} \right)
 = 0.0064 \theta_8 \nonumber
\end{eqnarray}
(Eqs.\ (62) 
and Appendix F
of OSV \cite{OSV}; cf. also 
Appendix A of
ED'A \cite{dambrosio}).
Here $g=0.952$ is our 
fiducial 
arm cavity's g-factor and 
$b=\sqrt{\lambda L/2\pi} = 2.603$ cm is
its transverse diffraction scale. 

\begin{figure}
 \includegraphics[width=3.in]{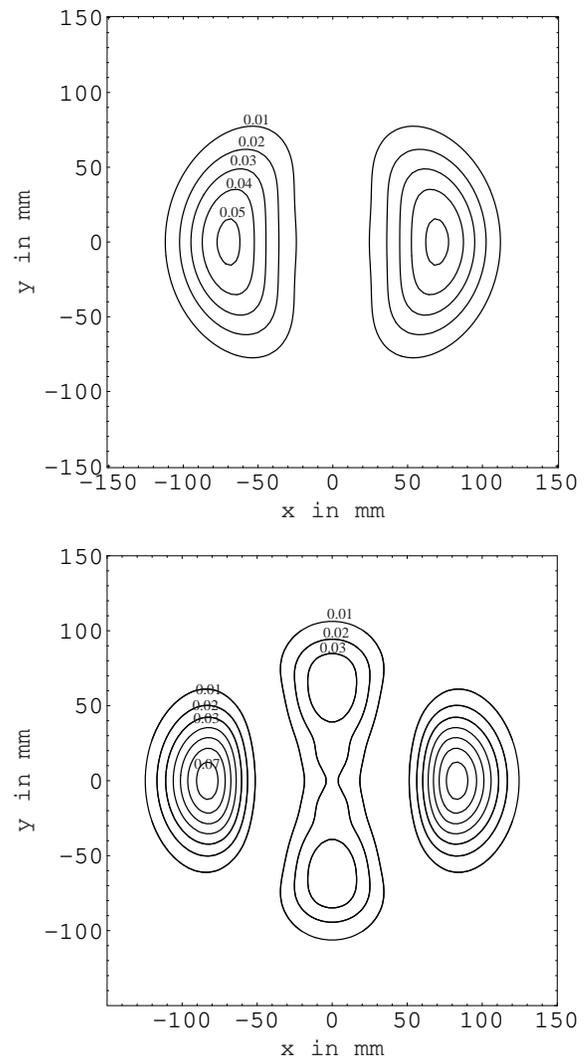}%
 \caption{Contour-diagram maps of the squared moduli $|u_1(\vec r)|^2$ (top) and
$|u_2(\vec r)|^2$ (bottom) of the parasitic modes excited by tilt of the ETM of a MH
arm cavity.  These maps were generated by O'Shaughnessy \cite{OSV}
using the eigenequation for a cavity with
infinite-radius MH mirrors, together with first- and second-order perturbation theory.
The axis about which the ETM is tilted (rotated) is
chosen to be vertical (the $y$ axis).  
The units in which $|u_j(\vec r)|^2$ is measured are
$b^{-2} = (2.603 \text{cm})^{-2}$; the peak values are $0.0526$ for $|u_1|^2$, and
$0.0726$ for the peak of $|u_2|^2$ on the $x$-axis and $0.0398$ for the peak
on the $y$-axis.  The contours are separated by 0.01 in units of $b^{-2}$; the
outermost contour in all cases is $0.01 b^{-2}$.  
See also Figs 15 and 16
of \cite{dambrosio} for maps of these same modes as computed by ED'A with
her FFT code.
\label{fig:u12R}}
 \end{figure}

For the MH cavity, $u_0$ is the mesa mode shown (for slightly different
beam radius) in Fig.\ \ref{fig:MHGaussModes}; and maps of 
$u_1$, and $u_2$, as computed
by O'Shaughnessy \cite{OSV} by solving the cavity eigenequation, are shown in 
Fig.\ \ref{fig:u12R}.  D'Ambrosio has used her
FFT code to compute maps that agree well with these
(Figs.\ 15 and 16 of \cite{dambrosio}).
The MH mixing
coefficients are\footnote{RO'S 
and ED'A both find $\alpha_1^{\rm MH} = 
0.0227 \theta_8$; for $\alpha_2^{\rm MH}$ RO'S finds $0.00018\theta_8^2$
and ED'A finds $0.00020\theta_8^2$.
} 
\begin{equation}
\alpha_1^\text{MH} = 0.0227 \theta_8\;, \quad \alpha_2^\text{MH} = 0.00019
\theta_8^2
\label{alphanMH}
\end{equation}
(Eqs. (63) of OSV \cite{OSV}; 
first equation on page 8 of ED'A \cite{dambrosio}).
Comparison of Eqs.\ (\ref{alpha1BL}) and (\ref{alphanMH}) shows that, 
\emph{to keep
the dominant, dipolar mode-mixing coefficient $\alpha_1$
at the same level in the 
fiducial
MH cavity as
in the 
fiducial 
spherical-mirrored cavity requires controlling the 
MH mirrors' tilt angle 
$3.5 \simeq 4$ 
times
more tightly.} 

[Note: If, instead of comparing our fiducial arm cavities (Sec.\ \ref{sec:fiducialconfigs}),
we had compared the baseline spherical-mirrored cavity with a MH-mirrored cavity that
has the same diffraction losses, we would have obtained $\alpha_1^{\rm sph} = 0.0047 \theta_8$ 
(instead of $0.0064 \theta_8$), and\footnote{We thank Pavlin Savov for computing this number for us.} $\alpha_1^{\rm MH} = 0.013\theta_8$ (instead of $0.0227\theta_8$).
We would thereby have concluded that, to keep the dipolar mode-mixing at the same level
in the MH cavity as in the spherical-mirrored cavity requires controlling the MH mirrors'
tilt angle $0.013/0.0047 = 2.8 \simeq 3$ times more tightly than the spherical mirror's tilt (rather than $0.0227/0.0064 = 3.5 \simeq 4$
times more tightly).  This illustrates the fact that the comparison of our fiducial
configurations gives approximately the same answers as would a comparison that focuses on
the baseline advanced-LIGO configuration.]

For our fiducial configurations, the
fractional power in the dominant, dipolar parasitic mode 
is 
$\alpha_1^2$, which will be doubled to 
\begin{eqnarray}
P_1^{\hbox{sph arm total}} &\simeq& 2(\alpha_1^\text{sph})^2\simeq0.001 
(\theta_8/3.5)^2 \;, \nonumber\\
P_1^{\hbox{MH arm total}} &\simeq& 2(\alpha_1^\text{MH})^2\simeq0.001 
\theta_8^2\;,
\end{eqnarray}
when the
ITM and ETM are both tilted but about uncorrelated axes.  This \emph{suggests
that, so far as the arm cavities are concerned (i.e., ignoring issues of
tilt in the recycling cavities), the tilt of MH mirrors need not be controlled much better
than $\simeq 10^{-8}$ rad.}  We have verified this by examining the effects
of the mode mixing on various cavity and interferometer parameters;
see the next three subsections.

\medskip

\subsubsection{Diffraction Losses}
\label{sec:DiffractionLosses}

One of us (ED'A), from her FFT simulations,
has estimated the influence of ETM tilt on diffraction losses to be 
(last equation on page 9 of \cite{dambrosio})
\begin{equation}
\mathcal L^{\prime\,\text{\;MH}}_0 = \mathcal L^\text{MH}_0(1+0.004\theta_8^2)\;,
\label{Diff}
\end{equation}
where $\mathcal L^{\text{MH}}_0 = 18$ ppm for our fiducial
mirrors.
This result has been confirmed to
a factor $\sim 2$ by RO'S  by combining the clipping approximation
with perturbation theory of
the cavity's eigenequation; Eq.\ (65) of \cite{OSV}.
The influence of ITM tilt should be about the same, thus doubling the coefficient
of $\theta_8^2$.
\emph{This tilt-induced increase in losses is so small that it can be ignored for
tilt angles below $\sim10^{-8}$ rad.}

\subsubsection{Arm Cavity Gain} 

We have computed 
(ED'A via the FFT code and RO'S via perturbation theory)
the following influence of ETM tilt on the arm
cavity gain\footnote{RO'S finds $G_\text{cav}^\text{MH} = 737(1-0.00055
\theta_8^2)$; ED'A finds $G_\text{cav}^\text{MH} = 746(1-0.00059  
\theta_8^2)$.
}
\begin{equation}
G_\text{cav}^\text{MH} = 740(1-0.00057 \theta_8^2)\;
\label{Gcav}
\end{equation}
(Eq.\ (68) of \cite{OSV}; 
Sec.\ III$\,$B of \cite{dambrosio}).  
This result assumes the baseline values for the power transmissivities of 
the ITM and ETM, 
and assumes for simplicity
that the only losses are diffraction losses; Sec.\ \ref{sec:rtIrtE}.
The factor 740 assumes
the cavity is driven by its best-fit Gaussian beam and thus is smaller by 
about $\gamma_0^2 =0.940$
than the gain in the untilted, 
fiducial
spherical-mirrored cavity.  The tilt of the ITM should produce
about the same
gain reduction as that of the ETM, thus doubling the coefficient of $\theta_8^2$ to $\sim 0.
001$.
This coefficient is small enough that  \emph{the tilt-induced decrease of
MH arm-cavity gain will be
negligible if $\theta$ is controlled to $\sim 10^{-8}$ rad.}

\subsubsection{Dark Port Power}

We have computed the influence of the tilt of one ETM on the dark-port output light (ED'A using
the FFT code and RO'S using perturbation theory).  Multiplying that result by four to deal 
with the case of all four cavity mirrors being tilted about uncorrelated axes, we find for
the fraction of the interferometer's input
power that winds up at the dark port in the fundamental mode $u_0$ and the parasitic modes
$u_1$ and $u_2$:
\begin{eqnarray}
P_0^\text{MH\; DP\; total} &\simeq& 1.0 \theta_8^4 \hbox{ ppm}\;,\nonumber\\
P_1^\text{MH\; DP\; total} &=&4 \gamma_0^2( \alpha_1^\text{MH})^2 \simeq 2000\theta_8^2 \hbox{ ppm}\;,\nonumber\\
P_2^\text{MH\; DP\; total} &=& 4 \gamma_0^2( \alpha_2^\text{MH})^2\simeq0.1\theta_8^4 \hbox{ ppm}\;.
\label{Pntotal}
\end{eqnarray}
(Eqs.\ (70) of \cite{OSV}
multiplied by 4; last four equations on page 8 of \cite{dambrosio}\footnote{In 
\cite{dambrosio} $P_{\rm DP}/P_0$ is half our $P_1^\text{MH\; DP\; total}$,
and $P_{\rm DP}^{\rm nondip}$ is half our $(P_0^\text{MH\; DP\; total} +
P_2^\text{MH\; DP\; total})$.
} multiplied by 2).
\emph{Without
an output mode cleaner, the dark-port power would primarily be in the dipolar mode $u_1$, and for 
$\theta < 10^{-8}$ rad it would constitute $< 0.2$ per cent of the input light.  The planned output mode cleaner will
wipe out this $u_1$ power and the power in mode $u_2$, leaving only the tiny 
fundamental-mode power, which should be totally negligible for 
$\theta$ below $10^{-8}$ rad.}

For comparison, the dark-port powers with the 
fiducial 
Gaussian beams are
\begin{eqnarray}
P_0^\text{sph\; DP\; total} &=&4(\alpha_1^\text{sph})^4 \simeq1.0 (\theta_8/3.5)^4 \hbox{ ppm}\;,\nonumber\\
P_1^\text{sph\; DP\; total} &=& 4(\alpha_1^\text{sph})^2\simeq 2000(\theta_8/3.5)^2 \hbox{ ppm}\;, \nonumber\\
\label{PnBLtotal}
\end{eqnarray}
which shows once again that the 
fiducial 
spherical-mirrored
arm cavities are $\sim 4$ times less sensitive to
tilt than the 
fiducial
MH arm cavities.

\subsection{Transverse Displacement of Arm Cavities' Mirrors}
\label{sec:Displacement}

When the ETM is displaced transversely through a distance $s$, the
cavity's fundamental mode gets changed from $u_0(r)$ to 
\begin{equation}
u'_0(\vec r) = (1-\zeta_1^2/2)u_0(r) + \zeta_1 w_1(\vec r) + 
\zeta_2 w_2(\vec r)\;,
\label{uprimedisplace}
\end{equation}
where the parasitic modes $w_n$, like the $u_n$'s, have unit norm,
$\langle w_n | w_n\rangle = \int |w_n|^2 d\text{Area} = 1$,
and have phase adjusted so the coupling coefficients $\zeta_n$ are real,
and where $\zeta_n \propto s^n$. 

R'OS has computed the coupling coefficients $\zeta_n$ for 
fiducial 
spherical
mirrors and 
fiducial
MH mirrors by applying perturbation theory to the cavity's
eigenequation;
and for MH mirrors, ED'A (unpublished)
has computed the influence of the ETM displacement 
on the arm cavity's gain using her FFT code, and 
has then deduced $\zeta_1^{\rm MH}$ by fitting  
to an analog of Eq.\ (68) of \cite{OSV}
in the limit of small displacements.
For both spherical and MH mirrors,
$\zeta_2$ is negligible compared
to $\zeta_1$ when the displacement is $s\ll b  \equiv \sqrt{\lambda L/2\pi}
= 2.60$ cm, so we shall ignore $\zeta_2$.  From the R'OS and ED'A
computations, we deduce
that the fiducial MH-mirrored cavities and fiducial spherical-mirrored
cavities are approximately equally sensitive to transverse displacements
of the ETM;  
their coupling coefficients 
are:\footnote{Here $\zeta_1^{\rm sph}$ 
and $\zeta_1^{\rm MH}$ 
are from R'OS \cite{OSV} Eqs.\ (57) and (58).
ED'A (unpublished) obtains $\zeta_1^{\rm MH} \simeq 0.008 s_\text{mm}$, 20
per cent smaller than R'OS; we regard her $\zeta_1^{\rm MH}$ as less reliable
than R'OS's because the phase of a parameter $\gamma_2$ used in her fit
is not known with sufficient reliability.
}
\begin{eqnarray}
\zeta_1^{\rm sph} &=& \left(\frac{(1-g)^{1/4}}{\sqrt2(1+g)^{3/4}}\right) {\frac{s}{b}}
= 0.008 s_\text{mm}\;,\nonumber\\
\zeta_1^\text{MH} &=& 0.010 s_\text{mm}\;.
\label{zetan}
\end{eqnarray}
Here $g=0.952$ is the 
fiducial 
arm cavity's g-factor, and 
$s_\text{mm}$ is the ETM's transverse displacement in millimeters.  The
MH cavity's lesser sensitivity to displacement presumably arises from
the flatness of its mirrors in the central region, where most of the
light power resides.   

The corresponding fraction of the arm-cavity carrier power driven into the (dipolar)
parasitic field $w_1$ is (cf.\ Eq.\ (59) of \cite{OSV})
\begin{equation}
P_1^{\hbox{arm total}} = \zeta_1^2 \simeq \left\{
\genfrac{}{}{0pt}{}{  100 (s/1.3 \text{mm})^2 \text{ppm}  \quad  \hbox{spherical,}} { 100 (s/1.0\text{mm})^2 \text{ppm} \quad   \hbox{MH.}\quad \quad} \right.
\end{equation}

The fraction of the input carrier power driven out the dark port when the
ETMs of both arm cavities are displaced through a distance $s$ but in
uncorrelated directions is about twice the above 
[cf.\  \cite{OSV}, Eq.\ (60)
multiplied by two, since here both ETMs are displaced and there just one]:
\begin{equation}
P_1^{\hbox{DP total}} = 2 \gamma_0^2 \zeta_1^2 
\simeq \left\{
{190 (s/1.3\text{ mm})^2 \text{ppm}  \quad \hbox{spherical,} \atop
 190 (s/1.0\text{ mm})^2 \text{ppm} \quad \hbox{MH.}\quad\quad} \right. 
\label{P1displacement}
\end{equation}

These 
coupling coefficients and parasitic-mode powers
are sufficiently small that  transverse
displacements are not a serious issue, and so shall ignore them in the rest of
this paper.  In any event, the low sensitivity to a change from 
fiducial 
spherical
to 
fidcuial
MH mirrors makes displacement a non-issue in the any decision about 
whether to use MH mirrors.

\subsection{Errors in the Arm Cavities' Mirror Figures}
\label{sec:FigureErrors}

\subsubsection{Billingsley's Worst-Case Figure Error}

Garilynn Billingsley (of the LIGO Laboratory, Caltech)
has provided us with a map of a worst-case figure
error, 
$\delta z_\text{wc}(x,y)$ [height error as function of Cartesian coordinates
in the transverse plane], produced by current technologies.  
Her map is based on the measured
deviation of a LIGO-I beam-splitter substrate from flatness.  
The measured
substrate had diameter 25 cm;  she stretched its deviation from flatness
(its ``figure map'')
to the 
fiducial mirror diameter of 32  cm, 
fit Zernike polynomials to the stretched
map, and smoothed the map by keeping only the lowest 36 Zernikes.

\begin{figure}
\includegraphics[width=3.4in]{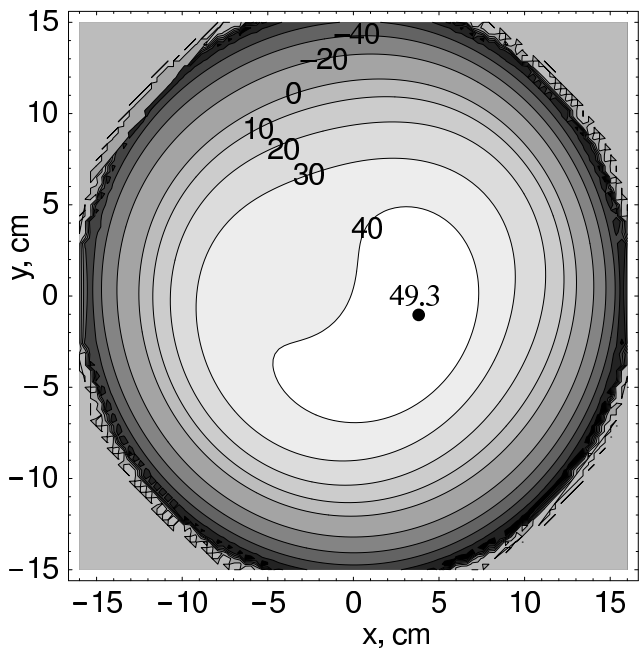}%
\caption{Contour diagram of Billingsley's worst-case figure error [height
$\delta z_\text{wc}$
in nanometers as a function of transverse Cartesian coordinates 
$(x,y)$  in centimeters].  The
hash at the outer edge of the mirror is an artifact of our numerical
manipulation of Billingsley's map.
}
\label{fig:GariMap}
\end{figure}

 We show a
contour diagram of the resulting figure map (figure ``error'')  in Fig.\ \ref{fig:GariMap}. 
In the central region (innermost 10 cm in radius), the peak to valley
error $\Delta z$ is about 30 nm, while in the outer region 
(10 cm to 16 cm in radius), it
is about 110 nm.  Billingsley thinks it likely that in the central region
(which dominates our considerations), 
peak-to-valley errors of $\Delta z \sim 5$ nm may be achievable for MH mirrors --- about 1/5
as large as in Fig.\ \ref{fig:GariMap};
and we have found that  $\Delta z = 0.2\times 30 = 6$ nm is small enough that the influences of the figure error scale, for $\Delta z\alt 6$ nm, as $\Delta z$ or $\Delta z^2$ with
higher-order terms producing 
$\alt 10$ per cent corrections.
Accordingly, in the analyses
described below we shall use Billingsley's map, scaled down in height
by a factor $\varepsilon$: 
\begin{equation}
\delta z = \varepsilon \delta z_\text{wc} (x,y)\;,
\label{epsilonDef}
\end{equation}
and we shall use $\varepsilon= 0.2$ and $\Delta z = 6$ nm as our fiducial values for $\varepsilon$
and $\Delta z$.
Jean Marie Mackowski (an expert in coating mirrors)
believes that $\Delta z \sim 2$ nm errors may be achievable for MH mirrors, if
the mirror figure is produced by coating; this corresponds 
to $\varepsilon \sim 0.07$.  For comparison, the figure errors for 
the spherical LIGO-I mirrors
are $\Delta z \sim 3$ to 6 nm (rms errors $\sim 1$ to 2 nm).

\subsubsection{Mode Mixing by Figure Errors Without Compensating Tilt}

We have computed the mode mixing in an MH cavity
when Billingsley's worst-case figure error, reduced by $\varepsilon \alt 0.2$, 
is placed
on the ETM.  As in our tilt studies, the computation was done independently
by ED'A using the FFT code and by RO'S using arm-cavity perturbation theory.

\begin{figure}
\includegraphics[width=3.4in]{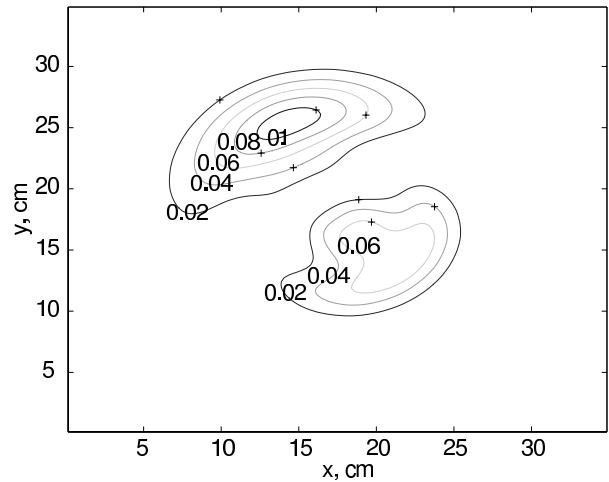}%
\caption{
The power distribution $|u'_0 - u_0|^2 = |\beta_1 v_1|^2$ 
(in units 1/m$^2$) 
of the deformation-induced
parasitic mode when the deformation $\varepsilon \delta z_\text{wc} (x,y)$ 
with $\varepsilon = 0.2$
is applied to the ETM of an MH arm cavity.  This map was computed by ED'A \cite{dambrosio}
using the FFT code; the map computed by RO'S using perturbation
theory \cite{OSV} is in reasonable agreement with this one (e.g., the heights of the two peaks
are $\{0.111, 0.080\}$ in the FFT map, and $\{0.110,0.088\}$ in the perturbation 
map).  The parasitic power $|\beta_1 v_1|^2$
scales as $(\varepsilon/0.2)^2$. }
\label{fig:DeformPowerNoTilt}
\end{figure}

By analogy with Eq.\ (\ref{u0prime}), the fundamental mode with deformed ETM
has the following form
\begin{equation}
u'_0 = (1-\beta_1^2/2)u_0 + \beta_1 v_1\;,
\label{uDeformed}
\end{equation}
where the parasitic mode $v_1$, like the $u_n$'s, has unit norm 
$\langle v_1 | v_1 \rangle = \int |v_1|^2 d\text{Area} = 1$, and has its phase adjusted so that
$\beta_1 \propto \varepsilon/0.2$ is real.
By contrast
with the tilt-induced mode mixing, where $u_1$ is 
dipolar (angular dependence
$\cos\varphi$),
the deformation parasite $v_1$ has a complicated
shape that depends on the details
of the deformation and that therefore contains a number of multipoles.
A map of the power distribution  $|\beta_1 v_1|^2$ of the admixed mode is shown
in Fig.\ \ref{fig:DeformPowerNoTilt}.

The fraction of the arm cavity power in the parasitic mode is
\begin{equation}
P_1^{\hbox{MH arm}} = \beta_1^2 =0.0011 (\varepsilon/0.2)^2\;
\label{PDeform}
\end{equation}
(Eqs.\ (73) of \cite{OSV}; 
first term in first equation on page 10, and also first equation on page
11, of \cite{dambrosio}).
The fraction of the interferometer's input power that goes out the dark port 
(if the interferometer is driven by the best-fit
Gaussian mode $u_d$ and if only one of the arm mirrors --- one ETM --- is deformed)
is given by
\begin{equation}
P_1^\text{MH\;DP} = \gamma_0^2 \beta_1^2 = 0.0010 (\varepsilon/0.2)^2
\label{PDeformDP}
\end{equation}
(Eq. (75) of \cite{OSV}; cf. also \cite{dambrosio}).
At the leading, $\varepsilon^2$, order in the deformation (the order to which we have
computed), this dark-port power is entirely in the
parasitic mode $v_1$.  Our FFT and perturbation-theory calculations agree on
the parasitic powers (\ref{PDeform}) and (\ref{PDeformDP}) to within 
about five per cent.

The dark-port power (\ref{PDeformDP}) and parasitic arm-cavity power (\ref{PDeform}) are
influenced primarily by the figure error in the central (10 cm radius) region of the
ETM, because about 96 per cent of the mesa-mode power is contained in that central region,
and only about 4 per cent in the outer region --- and of the outer 4 per cent, 3/4 (3 per cent) is 
in the annulus between 10 and 11 cm.  The insensitivity to outer-region deformations 
is fortunate, because Billingsley tells us that it will be much
easier to keep the figure errors small in the central region than in the outer region.

[We have verified 
the insensitivity to the outer-region deformations by evaluating (R'OS via
perturbation theory and ED'A via FFT simulations)
the dark-port power for a mirror deformation of the form
\begin{equation}
\delta z = \varepsilon_c \delta z_\text{wc}^\text{central} + \varepsilon_o
\delta z_\text{wc}^\text{outer}\;.
\end{equation}
Here
$\delta z_\text{wc}^\text{central}$ is equal to $\delta z_\text{wc}$ at $r<9.6$ cm
and is zero at $r>12.2$ cm, and between $9.6$ and $12.2$ cm,  
$\delta z_\text{wc}^\text{central}/\delta z_\text{wc}$ falls linearly from 1 to zero;
and similarly 
$\delta z_\text{wc}^\text{outer}$ is equal to $\delta z_\text{wc}$ at $r>12.2$ cm
and is zero at $r<9.6$ cm, and between $9.6$ and $12.2$ cm
$\delta z_\text{wc}^\text{outer}/\delta z_\text{wc}$ grows linearly from 0 to 1.
We find,
as a function of the central-region and outer-region weightings,\footnote{
The 
second coefficient, $(\epsilon_o/0.7)^2$, is from the ED'A FFT
calculations: a fit to numbers in the first and third equations on page
11 of \cite{dambrosio}, in which fit we (arbitrarily) ignored the possibility
of a cross term between $\epsilon_o$ and $\epsilon_c$.  
The RO'S perturbation-theory calcuations use mirrors of infinite radius,
and the modes admixed by the outer-region perturbations have sufficient
amplitude outside the radius of our fiducial mirror, $r>16$ cm, as to
make his results for the dependence on $\epsilon_o$ somewhat unreliable.
(The unrelieability arises from two effects: his admixed modes are wrong due 
to omitting the influence of the mirror edge, and his analysis misses the
influence of the mirror errors on diffraction losses.)
His analog of Eq.\ (\ref{P1outin}) is $P_1^{\hbox{MH DP}} 
\simeq 0.0010 [ (\epsilon_c/0.2)^2
+ \epsilon_o(\epsilon_o-\epsilon_c)/(0.7)^2]$, implying 5 times weaker
sensitivity to outer-region perturbations than central-region, by
contrast with the 3.5 times weaker sensitivity in ED'A's more reliable Eq.\ (\ref{P1outin}). 
} 
\begin{equation}
P_1^{\hbox{MH DP}} = \gamma_0^2 \beta_1^2 \simeq 0.0010 [ (\epsilon_c/0.2)^2
+ (\epsilon_o/0.7)^2]\;;
\label{P1outin}
\end{equation}
so that, for example, if Billingsley's worst-case perturbations are reduced
by $\epsilon_c = 0.2$ in the central region (to $\Delta z = 6$ nm), but are
are reduced only to $\epsilon_o = 0.7$ in the outer region
(so $\Delta z = 21$ nm there), the outer region will contribute about
as much power to the dark port as the inner region.]

When all four arm-cavity mirrors are subjected to uncorrelated deformations,
the arm-cavity parasitic power (\ref{PDeform}) will be increased by a factor 2 and the 
dark-port power (\ref{PDeformDP}) by a factor 4, to 
\begin{eqnarray}
P_1^\text{MH\; arm\; total} &=&2\beta_1^2 \simeq 0.0025 (\Delta z / 6 \text{nm})^2\;,\nonumber\\
P_1^\text{MH\; DP\; total} &=&4\gamma_0^2\beta_1^2\simeq 0.005 (\Delta z / 6 \text{nm})^2\
\label{PDeformFinal}
\end{eqnarray}
where $\Delta z$ is the peak-to-valley 
mirror deformation in the central region.  This 
suggests that, so far as arm-cavity mode mixing is concerned, peak-to-valley
figure errors of order 6 nm 
in the inner 10 cm
are acceptable.

\subsubsection{Mode Mixing by Figure Errors With Compensating Tilt}
\label{sec:MixWithTilt}

\begin{figure}
\includegraphics[width=3.5in]{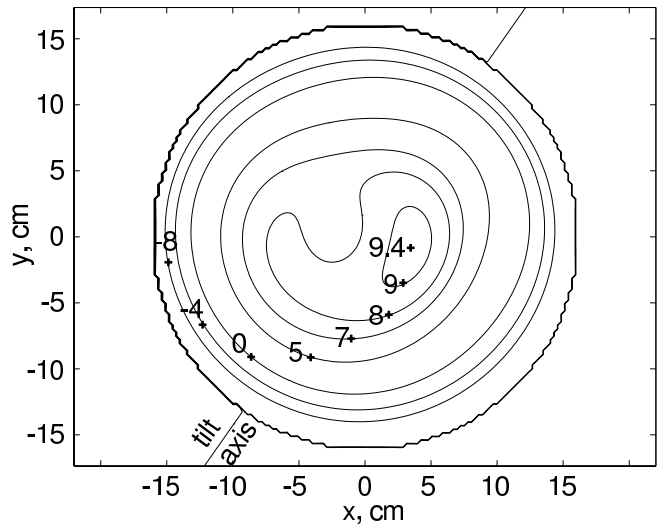}%
\caption{Billingsley's worst-case figure error 
reduced by a factor $\epsilon=0.2$,
when
the mirror is tilted to optimally reduce the mesa beam's odd-parity power:
$\delta z = 0.2 \delta z_\text{wc}+\theta_\text{compensate} \sin(\varphi-55^\text{o})]$.
The numbers on the contours are height in nanometers.
}
\label{fig:GariMapTilt}
\end{figure}

\begin{figure}
\includegraphics[width=3.0in]{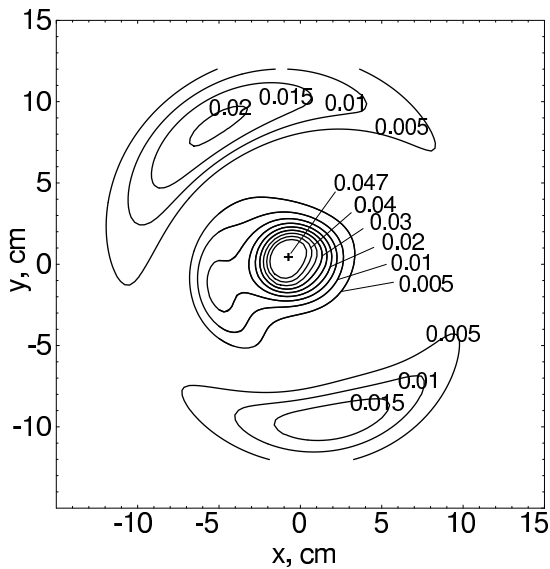}%
\caption{
The power distribution $|u'_0 - u_0|^2 = |\beta_{1c} v_{1c}|^2$ 
(in units 1/m$^2$) 
of the deformation-induced
parasitic mode when the tilt-adjusted
deformation $\delta z = 0.2 \delta z_\text{wc}+\theta_\text{compensate} \sin(\varphi-55^\text{o})]$ 
is applied to the ETM of an MH arm cavity.  This map was computed by RO'S
by applying perturbation theory to the cavity's eigenmode \cite{OSV}.  The map computed by
ED'A using the FFT code \cite{dambrosio} agrees reasonably well.  
}
\label{fig:DeformPowerTilt}
\end{figure}

The parasitic mode $v_1$ 
(Fig.\ \ref{fig:DeformPowerNoTilt}) 
contains a significant amount of dipolar
field, as one sees from the asymmetry of the map.  The advanced LIGO tilt control system, based
on a quadrant-diode readout of asymmetry in the power distribution $u'_0$, will tilt the
mirror so as to remove the overlap between the deformed parasitic field $v_1$ and
the dipolar-tilt parasitic field $u_1$.  
ED'A and R'OS have independently computed 
that the optimal tilt  is  about
$\theta_\text{compensate} = 1.3
\times 10^{-8} (\varepsilon/0.2)$ 
radians about a line rotated 55 degrees from the $x$ axis 
(Eqs.\ (77) of \cite{OSV}),
and have computed the resulting field
$u'_0 = u_0 + \beta_1 v_1+ \alpha_1 u_1$ with minimum parasitic-mode power.  Figure
\ref{fig:GariMapTilt} shows the mirror deformation after tilt, and Fig.\
\ref{fig:DeformPowerTilt} shows the parasitic power distribution $| \beta_1 v_1+ \alpha_1 u_1|^2$
for $\varepsilon=0.2$.  Notice that the tilt has largely but not completely removed the dipolar
asymmetry.  Some residual dipolar field remains --- that portion 
whose radial distribution prevents it from being 
compensated
by a tilt.  

We denote by $v_{1c}$ the tilt-compensated parasitic mode and by $\beta_{1c}$ its (real) amplitude,
so $\beta_{1c} v_{1c} = \beta_1 v_1 + \alpha_1 u_1$.  Then the cavity's eigenmode, with tilt 
compensation (including the second-order loss of power from $u_0$) is 
\begin{equation}
u'_0 = (1-\beta_{1c}^2/2)u_0 + \beta_{1c}v_{1c}\;,
\label{uDeformedCompensated}
\end{equation}
and optimization of the tilt compensation gives for the fraction of the power in the parasitic mode
\begin{equation}
P_{1c}^\text{MH\;arm} = \beta_{1c}^2 = 0.00040(\varepsilon/0.2)^2 = 0.00040 (\Delta z/6\text{nm})^2\;
\label{beta1c}
\end{equation}
(Eq.\ (79) of \cite{OSV}; last equation on page 10 of \cite{dambrosio},
divided by $\gamma_0^2 = 0.940$).
The fraction of the interferometer's input power that goes out the dark port (all in the parasitic modes), after this tilt compensation, is
\begin{equation}
P_1^\text{MH\;DP} =  \gamma_0^2 \beta_{1c}^2 = 0.00038 (\varepsilon/0.2)^2\;
\label{PDeformTiltDP}
\end{equation}
(Eq.\ (80) of \cite{OSV}; last equation on page 10 of \cite{dambrosio}).
Comparing with Eq.\ (\ref{PDeformDP}), we see that \emph{the figure-error-induced power
out the dark port is reduced by a factor $3$ by the compensating tilt of the deformed mirror.}

The dark-port power (\ref{PDeformTiltDP}) is for an interferometer with one deformed
MH arm mirror.  When all four mirrors are deformed in uncorrelated ways, 
the arm-cavity parasitic power will
be doubled and the dark-port power will be quadrupled:
\begin{eqnarray}
P_1^\text{MH\; arm\;total} &=& 2\beta_{1c}^2 \simeq 0.0008 (\Delta z / 6 \text{nm})^2\;,
\nonumber\\  
P_1^\text{MH\; DP\;total} &=& 4\gamma_0^2\beta_{1c}^2 \simeq 0.0015 (\Delta z / 6 \text{nm})^2\;.  
\label{PDeformTiltFinal}
\end{eqnarray}
{\em This suggests that, so far as arm-cavity mode mixing is concerned, 
we could live with central-region deformations as large as 10 nm.
Recycling-cavity issues, discussed below, will place much tighter constraints
on the mirror figures.}

We note in passing that the computations reported in this subsection have verified the ability of the 
LIGO control system to measure the lowest-radial-order dipolar component of a mirror figure error, 
and remove it by applying a tilt.

\subsubsection{Comparison with Mode Mixing for Baseline Spherical Mirrors with Tilt-Compensated Figure Errors}

As part of her modeling of the baseline configuration of advanced
LIGO (a different project from that reported here), Ed'A has evaluated 
tilt compensation and mode mixing for an interferometer with the baseline (BL) spherical mirrors,
 to which the figure error of Fig.\ \ref{fig:GariMap} with $\varepsilon=0.2$ has been applied.  For the BL spherical mirrors, 95 per cent of the light power is contained inside a circle with radius
$r_{\rm 95\%} \simeq 7.3$ cm, by contrast with the fiducial MH mirrors, where $r_{\rm 95\%} \simeq 10$ cm (and the fiducial spherical mirrors with $r_{95\%} \simeq 8$ cm).  Since $(7.3/10)^2 \simeq 0.5$, the BL Gaussian beam feels a central region of the deformed mirror with area approximately
half that felt by the fiducial mesa beam.   

Interestingly, the tilt compensation that removes the lowest-radial-order dipolar deformation is
twice as large for the BL spherical case as for the fiducial MH case:  $\theta_{\rm compensate}^{\rm BL\;sph} \simeq 2.6\times 10^{-8} (\varepsilon/0.2)$ compared to 
$\theta_{\rm compensate}^{\rm fid\;MH} \simeq1.3\times 10^{-8} (\varepsilon/0.2)$.  The BL tilt is about a line
rotated 76 degrees from the $x$-axis, compared to 55 degrees for the
fiducial MH case.  These differences are due to the different regions of the mirror that the two
beams feel.

After tilt compensation,  the peak-to-valley figure-error
variations in the 95-per-cent-power region are [for figure errors reduced from worst case
by $\varepsilon=0.2$] $(\Delta z)^{\rm BL\;sph}_{95\%} \simeq 2$ nm (inner
7.3 cm), by contrast with $(\Delta z)^{\rm fid\; MH}_{95\%} \simeq 6$ nm (inner 10 cm).  This
factor 3 difference is caused by Billingsley's figure errors becoming larger rapidly 
with increasing radius ---
an issue that will have to be dealt with if MH mirrors are used in advanced LIGO.

D'Ambrosio's analysis of mode mixing by the tilt-compensated figure errors gives, for the 
fraction of arm-cavity power driven into parasitic modes, 
$P_{1}^\text{BL\;sph\;arm\,total} \simeq 0.0008 [(\Delta z)_{95\%}/12{\rm nm}]^2$, compared to 
$P_1^\text{MH\; arm\;total} \simeq 0.0008 [(\Delta z)_{95\%}/6{\rm nm}]^2$ [Eq.\ (\ref{PDeformTiltFinal})].
Thus, for the same amount of mode mixing, the BL spherical mirrors can have twice as large 
peak-to-valley figure errors as the fiducial MH mirrors, and in the BL spherical case the relevant
figure errors are confined to a region half as large in area ($r_{95\%} \simeq 7.3$ cm) as in the fiducial MH
case ($r_{95\%} \simeq 10$ cm).  Presumably, the two-fold greater sensitivity of MH arm cavities to figure errors is caused by the greater degeneracy of their modes (last paragraph of Sec.\   \ref{sec:ParasiticModes}).  If we had compared fiducial MH mirrors with fiducial spherical
mirrors instead of with BL spherical mirrors, the diffferences would have been more modest.

\subsection{Influence of Mirror Tilt and Figure Errors on Thermoelastic Noise}
\label{sec:DeformedTE}

When one MH mirror of an arm cavity  is given the deformation $\varepsilon \delta z_\text{wc}$,
 the resulting deformation of the arm-cavity power distribution, 
$\delta |u'_0|^2 = |u'_0|^2 - |u_0|^2$
increases
the thermoelastic noise.  The following argument (due to RO'S 
Secs.\ IV$\,$C$\,$4 and VII$\,$E$\,$4 of \cite{OSV}) 
shows that, 
at leading (linear) order in
$\varepsilon/0.2$, \emph{only the circularly symmetric portion of the parasitic mode $\beta_1 v_1 =
u'_0 - u_0$ contributes to the thermoelastic noise increase}:
The fractional noise increase is equal to the fractional
increase in the thermoelastic
noise integral $I = \int (\vec\nabla \Theta )^2 d\text{volume}$ 
[Eqs.\ (\ref{Sh}) and (\ref{IA})]: 
\begin{equation}
\frac{\delta S_h^{\rm TE\; MH}}{ S_h^{\rm TE\; MH}}
= \frac{\delta \bar I}{\bar I} \propto \int (\vec\nabla\Theta)\cdot(\vec\nabla\delta\Theta)
d{\rm volume}\;.
\end{equation}
Here $\Theta$ is not a tilt angle but rather is the expansion (fractional increase
of volume) of the substrate material when a static pressure $P \propto |u'_0|^2$ is applied to
the mirror face.  Since the unperturbed mesa beam is circularly symmetric, so will be the unperturbed expansion $\Theta$, which means that only
the circularly symmetric portion of the expansion perturbation $\delta\Theta$, and thence
the circularly symmetric portion of the pressure perturbation $\delta P \propto \delta |u'_0|^2$ 
will contribute
to the noise increase.  At leading (linear) order in the mirror deformation $\varepsilon/0.2$, the
circularly symmetric portion of $\delta|u'_0|^2$ arises solely from the circularly symmetric
portion of $\delta u'_0 = \beta_1 v_1$; thus, as claimed, only the circularly symmetric portion
of $\beta_1 v_1$ can increase the thermoelastic noise.

This same argument shows that \emph{the mesa-mode deformation produced by mirror tilt cannot
influence the thermoelastic noise at first order in the tilt angle}; and therefore, \emph{we need not
be concerned about the influence of tilt on the thermoelastic noise --- whether the tilt is
unintended, or is being used in a controlled way to compensate the errors in the mirror
figures.}

RO'S 
and SS have independently computed
the fractional increase in the thermoelastic noise integral at linear
order in $\varepsilon/0.2$; cf.\ Eq.\ (81) in OSV \cite{OSV}.  Their result, multiplied by four to account for four arm-cavity mirrors, is
\begin{equation}
{\delta S_h^\text{TE,\;MH\; total}\over S_h^\text{TE,\;MH}} = 0.14 (\varepsilon/0.2) \simeq 0.14 
 (\Delta z / 6 \text{nm})\;.
\label{TEDeformIncrease}
\end{equation}
\emph{This 14 per cent increase of $S_h$, when all four mirrors are subjected to 6 nm figure
errors in their central regions, is to be compared with the factor $1/0.34 =$ 295 per cent decrease
in thermoelastic noise achieved by switching from spherical mirrors to MH mirrors.}  There may
also be a small increase in thermoelastic noise when a spherical mirror is deformed. 
\emph{Assuming,
conservatively, no 
deformation-induced noise increase
for spherical mirrors, switching from spherical to 6nm-deformed MH mirrors will reduce
the thermoelastic noise 
by a factor $0.34\times1.14 = 0.39$, which in turn will
increase the distance for NS/NS binaries
from 364 Mpc to about 490 Mpc  (Fig.\ \ref{fig:Range}) and increase the event rate by about a factor 
2.45}; cf.\ Table \ref{tbl:NSRange}.

\subsection{Mesa vs.\ Gaussian Beams in Recycling Cavities} 
\label{sec:RecyclingCavities}

In LIGO-I interferometers, and in the baseline design for advanced LIGO interferometers,
the greatest sensitivity to mirror tilts and figure errors occurs in the power recycling (PR)
and signal recycling (SR) cavities.  This sensitivity arises from the near degeneracy of
the recycling cavities, which strongly enhances error-induced mode mixing.  One might worry 
that for MH mirrors,
with their greater central-region flatness, this severe mode mixing 
might
be made
substantially worse.

We have examined this question and conclude that 
\emph{for the two
wideband advanced LIGO interferometers, 
there is not much
difference between the 
fiducial 
Gaussian beams and the 
fiducial 
mesa beams, with
regard to their susceptibility to mode mixing in the recycling cavities}.  
The only 
significant difference arises from the fact that the mesa beams are larger and therefore
sample, with significant power, larger-radii regions of the mirrors (the regions between, say,
8 cm radius and 10 cm radius), where the deformations may be worse.

The reason that the mode mixing is only marginally sensitive to
the beam shape is quite simple:   
If heterodyne readout were used, then 
once RF modulated sideband light gets into the RF optical cavity bounded by
the power recycling (PR) mirror, the signal recycling (SR) mirror and the two
input test-mass (ITM) mirrors, the sideband light
would make roughly 
$\mathcal N_\text{RF} 
\sim \frac12($cavity finesse $\mathcal F_{PR})\simeq (\pi/2) /(1-R_\text{PR}) \simeq 25$ round trips before losing 95 per cent of its power out the
beam
splitter's dark port.\footnote{Our 
definition ${\mathcal N}=($number of round trips$) \simeq {\mathcal
F}/2$, based on 95 per cent power loss, differs by $4/\pi \simeq 1.27$
from a more conventional
${\mathcal N} = (2/\pi){\mathcal F}$ based on the phase shift of light
emerging from the cavity.   
}
(Here $R_\text{PR} = R_\text{SR} 
\simeq 0.94$ is the power reflectivity 
of the PR and SR mirrors; they are made equal so as to transmit the RF light with maximum
efficiency to the photodetector.)
And once signal light gets into the signal recycling cavity,
it makes roughly $\mathcal N_\text{SR}      
\sim \frac12 \pi/(1-\rho\sqrt{R}) \sim 40$ round trips before losing 95 per cent
of its power out the dark
port or back into the arm cavities. (Here $R=0.995$ is the ITM power 
reflectivity and $\rho = \sqrt{0.93}$ is the amplitude reflectivity of the
SR mirror, in the notation of Buonanno and Chen \cite{BuonannoChen3}.) 
The Fresnel length (transverse diffraction scale) for light
that makes $\mathcal N $ round trips in either recycling cavity, 
with cavity lengths
$\ell_{RF} \simeq 20$ m and $\ell_{SR} \simeq 10$ m,
is
\begin{equation}
r_F = \sqrt{\lambda_o 2\ell {\mathcal N}} \simeq 3 \text{cm}\;,
\label{rF}
\end{equation}
for both cavities; here
$\lambda_o = 1 \mu$m is the light's wavelength.
This Fresnel length
is 
$\sim 1/2$ of the $\sim 5$ cm scales on which the ideal mirror shapes and the
central-region worst-case mirror errors vary, and it is small compared to the
$\sim 15$ to 20 cm diameter beams themselves.  There thus is only modest
diffractive
coupling between light rays, and
\emph{the light bouncing back and forth in each recycling cavity
is describable, to moderately good accuracy, by geometric optics.} 
Moreover, because the mirrors
(whether MH or spherical) are nearly flat and nearly identical, the light's rays, to
rather good accuracy, are all parallel to the optic axis and to each other and are 
thus decoupled from each other.  If the mirrors and beam splitter were perfect and ideal
in shape, the extreme 
length of their radii of curvature, $\agt 50$ km, compared to the optical pathlength
in the recycling cavities,
$2 \ell \mathcal N  \simeq 1$ km, 
would guarantee
that the mesa beam would resonate equally well in the ideal MH-mirrored cavity or in the ideal
spherical-mirrored cavity, or in a precisely flat-mirrored cavity; and the 
fiducial 
Gaussian
beam would also resonate, equally well, in all these cavities.  

If 
an advanced LIGO
interferometer is operated in narrow-band mode, then the
number of round trips the signal light makes in the SR cavity will be much
larger than 40, and the geometric optics approximation will begin to fail
significantly.
More specifically, for 
ITM 
reflectivity $R=0.995$ and
optimized narrow banding at \{500 Hz, 1000 Hz\},
the SR mirror's amplitude reflectivities are $\rho = \{0.994, 0.9985\}$ 
[see discussion following Eq.\ (\ref{calA}) below], corresponding to
a number of round trips in the SR cavity $\mathcal N_\text{SR} \simeq \{180,\;
400\}$ and Fresnel lengths $r_F \simeq \{6 \text{cm},\; 9 \text{cm}\}$. These Fresnel
lengths (the transverse scale for diffractive light spreading) are  about 1/3 to 1/2 the
 95-percent-power diameter of the beam, 16 cm 
(fiducial 
Gaussian) and
20 cm 
(fiducial
mesa).  As we shall see below, this means that geometric optics can be
used to get a rough upper limit on the fractional 
decrease of signal strength
due
to tilt and irregularities of the SR cavity's mirrors.

\subsection{
Decrease of Signal Strength
Due to Mirror Tilts}
\label{sec:TiltShot}

\subsubsection{Foundations}

The mirror tilts produce a mismatch between various modes of the light, thereby 
decreasing the signal strength at the interferometer output (the beam splitter).  
We shall focus on 
the signal-strength decrease 
at the 
frequency
of the signal light's optical resonance in the arm cavity.  This optical resonance is the one that is used to produce a noise 
[$S_h(f)$] 
minimum
for a narrowbanded advanced LIGO interferometer, and it is the right-hand 
minimum (at $f \simeq 230$ Hz) of the optical noise for the standard wide-band advanced LIGO interferometer (Fig.\ \ref{fig:NoiseCurves}),
and approximately the minimum of the wide-band interferometer's total noise.
At this frequency, for MH mirrors with their reduced thermoelastic noise, the dominant noise source is
shot noise; for spherical mirrors, short noise and thermoelastic noise are comparable 
at this frequency, but in a small piece of our analysis, we shall idealize the shot noise as dominant.

The (unit-norm) modes whose mismatch 
decreases the signal strength
are the following: 
\begin{enumerate}
\item $u_0$, the eigenmode of perfect arm cavities.
\item $u'_0$, the carrier's eigenmode in an arm cavity with tilted ITM and ETM.
\item $u'_s$, the signal field's eigenmode in an arm cavity, 
at the center of its optical resonance, with tilted ITM, ETM, and SRM (signal recycling mirror).
\item $u'_t$, the field produced when $u_0$ is transmitted through the signal recycling
cavity with tilted ITM and SRM.
\item $u'_r$, the reference-light field that is beat against the signal light to produce the
input to the photodetector.  For the baseline homodyne readout system, this will be
carrier light $u'_0$ transmitted through the ITM and SR mirror to the photodiode.  If 
heterodyne readout were used, it would be side-band light transmitted through
the 
RF cavity 
to the photodetector.
\end{enumerate}

For each primed field $u'_\ell$, we denote by $\delta_\ell$  the fraction of its light power that is in 
parasitic modes and thus has been lost from 
the fundamental mode $u_0$ due to mirror tilt: 
\begin{equation}
\langle u_0, u'_\ell \rangle^2 = 1 - \delta _\ell\;.
\label{deltadef}
\end{equation}
 
The signal amplitude entering the photodetector is proportional to
\begin{equation}
S \propto \langle u'_r,\tilde \tau' u'_s\rangle \langle u'_s,u'_0 \rangle \langle u'_0,u_d \rangle\;.
\label{Sdef}
\end{equation}
The sequence of terms, from right to left, have the following meanings, and we approximate them
as follows:
\begin{enumerate}
\item
$\langle u'_0,u_d \rangle$ describes the influence of tilts on the
driving of the arm cavity's eigenmode by the Gaussian
driving field.  For simplicity, we neglect the tiny coupling of the driving field to the 
second-order parasitic mode $u_2$ contained in $u_0$ and therefore approximate this coupling 
amplitude by $\langle u'_0, u_d \rangle = \gamma_0 \langle u'_0, u_0 \rangle = \gamma_0 ( 1 - 
\alpha_{1E}^2/2 - \alpha_{1I}^2/2) = \gamma_0(1-\delta_0/2)$.  Here 
the subscripts $I$ and $E$ denote the contributions
from the tilts of the ITM and ETM. 

\item
$\langle u'_s,u'_0 \rangle$ describes the influence of tilts on the driving of the arm cavity's 
signal field by its carrier field (via the gravitational-wave-induced motion of the mirrors).  
For simplicity we neglect the 
(nonzero) overlap  
between the parasitic modes contained in 
$u'_s$ and $u'_0$, thereby obtaining $\langle u'_s,u'_0 \rangle = (1-\delta_s/2-\delta_0/2)$.

\item
$\langle u'_r,\tilde \tau' u'_s\rangle$ describes the influence of tilts on (i) the passage of the
signal $u'_s$, through the SR cavity (with cavity transmissivity $\tilde\tau'$ in the notation
of Buonanno and Chen \cite{BuonannoChen3}), and on (ii) the overlap of the transmitted
signal light with the reference
light to produce the photodetector current.  Again we neglect correlations between the parasitic
components of the fields and therefore approximate the influence of the tilts by
$\langle u'_r,\tilde \tau' u'_s\rangle \propto (1-\delta_r/2-\delta_s/2-\delta_t/2)$.
The $\delta_j$ terms represent the loss of overlap due to the parasitic-mode fields (assumed
uncorrelated) contained in $u'_r$ (the $\delta_r$ term), contained in $u'_s$ (the $\delta_s$ term), 
and generated by the passage of the signal light through the SR cavity, whose mirror tilts
deform the transmissivity $\tilde \tau'$,
(the $\delta_t$ term).
\end{enumerate}
 
If there is no mode cleaner on the interferometer output, then the rms amplitude of the shot noise 
(the dominant noise source at our chosen frequency) 
is 
$N\propto \sqrt{\langle u'_r,u'_r \rangle} = 1$;
i.e. the parasitic-mode components of $u'_r$ 
contribute to the rms noise amplitude along with the fundamental-mode component.  However, a mode 
cleaner will remove the parasitic components, so that 
$N\propto \langle u_0,u'_r \rangle = 
1-\delta_r/2$.

Combining the above approximations to the various terms, we find for the ratio of noise power
to signal power (which is proportional to the spectral density of 
noise $S_h$ referred to the gravitational-wave input $h$),
at
the minimum of the optical resonance:
\begin{equation}
S_h \propto {N^2\over |S|^2} \propto 1 + 2 \delta_0 + 2 \delta_s + \delta_t
+ \left\{ {\delta_r\;, \hbox{ no mode cleaner} \atop 0 \hbox{ with mode cleaner.} }\right.
\label{ShShot}
\end{equation} 
(Note that we have ignored the increase in shot noise due to carrier-light parasitic fields
going out the
dark port, Eqs.\ (\ref{Pntotal}) and (\ref{PDeformTiltFinal}), under the assumption that it is negligible, either because of an
output mode cleaner or because the arm-cavity-mirror figures and tilts are adequately
controlled.)

We shall now examine the various contributions to 
the increase of $S_h$ 
one by one.

\subsubsection{Carrier Light in Arm Cavity}
The fraction of the arm-cavity carrier power that is driven into parasitic
modes by tilts of the ETM and ITM is $\delta_0 = \alpha_{1E}^2 + \alpha_{1I}^2$.
The loss of this carrier power to parasites  
reduces the signal power $S^2$ and thence increases the gravitational-wave noise 
by
\begin{equation}
\left({\delta S_h\over S_h}\right)^\text{MH}_\text{carrier} =
2 \delta_0 = 4 \alpha_1^2 = 0.01 \left(\theta\over 2\times 10^{-8} \right)^2\;.
\label{dShCarrier}
\end{equation}
Here we have assumed that both mirrors are tilted through the 
same angle $\theta$ but about uncorrelated axes, 
we have assumed MH mirrors, and we have used Eq.\ (\ref{alphanMH}) for $\alpha_1$.  
The 
fiducial 
spherical mirrors are four
times less sensitive to tilt, so to keep  
this contribution to 
$S_h$
 below one per cent,
we must control the ITM and ETM tilts to an accuracy
\begin{equation}
\theta_{1\%}^\text{MH} = 2\times 10^{-8}\;, \quad
\theta_{1\%}^\text{sph} = 8\times 10^{-8}
\;.
\label{thetaCarrier}
\end{equation}
These are modest constraints on tilt.

\subsubsection{Signal Light in Arm Cavity}
\label{sec:SignalInArm}

The signal recycling (SR) cavity presents a complex amplitude reflectivity
$\tilde \rho' = e^{-\epsilon/F} e^{i\lambda/F}$ to the arm cavity's
signal light (Eqs.\ (5) and (13) of Buonanno and Chen \cite{BuonannoChen3}).   
Here $F=c/2L$ is the interferometer's free spectral range,
$\epsilon = \epsilon(R,\rho,\phi)$ and $\lambda = \lambda(R,\rho,\phi)$ are real functions of the ITM
power reflectivity $R$, the SRM amplitude reflectivity $\rho$ and
the SR cavity's tuning phase $\phi = (k \ell)_{\text{mod} \,2\pi}$, with $\ell$
the length of the cavity; and our notation is that of Buonanno and Chen
\cite{BuonannoChen3}.  Tilts of the ITM and SRM produce a spatially
variable reflectivity $\tilde\rho'$.  The spatial variations of the modulus
$e^{-\epsilon/F}$ of $\tilde\rho'$ presumably will have much less
influence on the arm cavity's signal eigenmode $u'_s$ than the spatial
variations of the phase.  (This claim deserves to be checked.)  Assuming
this is so, then the dominant influence of an ITM or SRM tilt $\theta$ is
to produce a spatially variable mirror displacement 
\begin{equation}
\delta z = \theta r \sin\varphi 
\label{deltazTilt}
\end{equation}
(where $\varphi$ is azimuthal
angle and $r$ is radius), which in turn 
(\emph{in the SR cavity's geometric optics limit})
produces a spatially variable
phase of the cavity reflectivity, $\text{arg}(\tilde\rho') = \delta\lambda/F
= (d\lambda/d\phi) (k/F) \delta z $.  If the cavity were replaced by a
single mirror that is displaced through a distance $\delta z_\text{eff}$,
then this phase change would be $2 k \delta z_\text{eff}$.  Correspondingly,
the tilt of the ITM or SRM produces an effective mirror displacement
$\delta z_\text{eff} = \mathcal A \delta z$, where the amplification factor 
$\mathcal A$ is given by
\begin{equation}
\mathcal A = {\delta z_\text{eff} \over \delta z} = {d\lambda /d\phi \over 2 F}
= (1-R)\rho {2\rho + (1+\rho^2) \cos2\phi \over (1+\rho^2) + 2 \rho\cos2\phi}\;;
\label{calA}
\end{equation}
see Eq.\ (18) of Buonanno and Chen \cite{BuonannoChen3}.

We shall focus on three configurations for the SR cavity: (i) The standard
wideband advanced LIGO configuration (denoted ``WB''), 
for which $R=0.995$, $\rho =
\sqrt{0.93}$, and $\phi = \pi/2 - 0.06$.  (ii) An interferometer narrowbanded
at a frequency $f= \lambda/2\pi \simeq 500$ Hz with bandwidth $\Delta f =
\epsilon/2\pi \simeq 50$ Hz, which has a noise minimum of $\simeq1\times 10^{-24}/\sqrt{\rm
Hz}$; this configuration (which we shall denote ``500'') is produced by
$R=0.995$, 
$\rho = 0.994$, and $\phi = 1.541$.  (iii) An interferometer
narrowbanded at $f=\lambda/2\pi \simeq 1000$ Hz with bandwidth $\Delta f =
\epsilon/2\pi \simeq50$ Hz (and so denoted ``1000''), which has a noise minimum of $\simeq1\times 10^{-24}/\sqrt{\rm
Hz}$ and parameters $R=0.995$, $\rho = 0.9985$, $\phi = 1.556$.  For these
three configurations the amplification factor is
\begin{equation}
\mathcal A_\text{WB} = 0.27\;, \quad
\mathcal A_{500} = 1.4\;, \quad
\mathcal A_{1000} = 5.7\;.
\label{calAvalues}
\end{equation}

We have chosen to compute the 
signal loss 
at the optical resonance
so the signal field $u'_s$ in the arm cavity will be on resonance, just as the
carrier field is.  This allows us to translate our carrier-field results over to the
signal field with only one change: the influence of the tilts of the SRM and ITM
must be multiplied by the amplification factor $\mathcal A$.  Therefore, the
fraction $\delta _s$ of the signal field's power that is in the tilt-induced parasitic
modes is $\delta_s = \alpha_{1E}^2 + \mathcal A^2 (\alpha_{1I}^2 + \alpha_{1 SR}^2)$.  
The influence $\alpha_{1E}^2$ of the ETM is the same as in the case of the carrier,
which we have already dealt with, so we shall ignore it here and focus on the
two mirrors that make up the SR cavity: the ITM and the SRM.  If they both
have the same tilt angles $\theta$ (but about uncorrelated axes) so $\alpha_{1I}^2 =
\alpha_{1SR}^2 \equiv \alpha_1^2$, then
these tilts produce a fractional 
loss of signal power and thence fractional increase in gravitational-wave noise given by
\begin{equation}
\left({\delta S_h\over S_h}\right)_\text{signal} =
2 \delta_s = 4\mathcal A^2 \alpha_1^2 \;.
\label{dShSignal}
\end{equation}
This is greater by the factor $\mathcal A^2$ than the noise (\ref{dShCarrier})
due to loss of carrier light into parasitic modes, and correspondingly to keep
this fractional increase of 
$S_h$
below one per cent requires controlling
the ITM and SRM tilts to an accuracy $1/\mathcal A$ of that in Eq.\ (\ref{thetaCarrier}):
\begin{eqnarray}
\theta_{1\%}^\text{MH\;WB} &=& 7\times 10^{-8}\;, \quad 
\theta_{1\%}^\text{sph\;WB} = 30\times 10^{-8}\;,\nonumber\\
\theta_{1\%}^\text{MH\;500} &\agt& 1.4\times 10^{-8}\;,\quad
\theta_{1\%}^\text{sph\;500} \agt 6\times 10^{-8}\;,  \nonumber\\
\theta_{1\%}^\text{MH\;1000} &\agt& 0.4\times 10^{-8}\;, \quad
\theta_{1\%}^\text{sph\;1000} \agt 1.4\times 10^{-8}\;.\nonumber\\
\label{thetaSignal}
\end{eqnarray}

For the narrowbanded interferometers these limits are only lower bounds
on $\theta_{1\%}$ because of the failure of the geometric optics limit.  
As we have seen, the Fresnel length for light trapped in the SR cavity
is about 1/2 to 1/3 of the 95-percent-power beam diameter, so 
transverse spreading of the light will reduce somewhat the SR 
cavity's amplification factor $\mathcal A$ and thence the influence of tilt on
the beam asymmetry.  We {\it guess} that this reduction might
increase $\theta_{1\%}^{1000}$ by a factor of
order 2  over the geometric-optics limit, Eq.\ (\ref{thetaSignal}); but since
$\mathcal A$ is only about 1 for 
$\theta_{1\%}^{500}$,
 we {\it guess}
that there is little increase in $\theta_{1\%}^{500}$; so
\begin{eqnarray}
\theta_{1\%}^\text{MH\;500} &\simeq& 1.4\times 10^{-8}\;,\quad
\theta_{1\%}^\text{sph\;500} \simeq 6\times 10^{-8}\;,  \nonumber\\
\theta_{1\%}^\text{MH\;1000} &\simeq& 0.8\times 10^{-8}\;, \quad
\theta_{1\%}^\text{sph\;1000}\simeq 3\times 10^{-8}\;.\nonumber\\
\label{thetaSignal1}
\end{eqnarray}

\subsubsection{Transmission of Signal Light Through SR Cavity}
\label{sec:TransmissionTilt}

When the ITM or SRM is tilted through an angle $\theta$, producing 
a spatially dependent mirror
displacement $\delta z = \theta r \sin\varphi $, it alters the SR cavity's
transmissivity by a spatially dependent amount $\delta \tilde \tau'
= (d\tilde\tau'/d\phi) k \delta z$, \emph{in the geometric optics limit}.  When an undistorted signal beam
$u_0$ passes through this spatially variable transmissivity, a fraction
\begin{equation}
\langle u_0, | \delta\tilde\tau'/\tilde\tau'|^2 u_0 \rangle 
= {1\over2}\mathcal B^2 k^2 \langle r^2\rangle \theta^2\;.
\label{deltatau}
\end{equation}
gets transferred to parasitic modes.
Here $k = 2\pi/\lambda_o$ is the wave number, $\langle r^2\rangle = \langle u_0,r^2 u_0\rangle$ 
is the beam's mean square radius, which has the values
\begin{eqnarray}
\label{rRMS}
\langle r^2 \rangle &=& (6.95\text{cm})^2 \quad \hbox{ for fiducial mesa beam}\\
\langle r^2 \rangle &=& r_o^2 = (4.70 \text{cm})^2  \quad \hbox{ for fiducial Gaussian beam},
\nonumber
\end{eqnarray}
and
\begin{equation}
\mathcal B^2 = \left | { d\tilde\tau'/d\phi \over \tilde\tau'} \right |^2 = {4 R\rho^2 \over
1+R\rho^2 + 2 \sqrt{R} \rho \cos2\phi}\;.
\label{calBDef}
\end{equation}
Here we have used Eq.\ (11) of Buonanno and Chen \cite{BuonannoChen3} for 
$\tilde\tau'$,
with the factor $e^{i\phi(\vec r)}$ in the numerator removed, so as to obtain the
transmissivity that carries the field from an input transverse plane to an 
output transverse plane in the presence of the mirror tilt (which gives $\phi$ its
dependence on transverse position $\vec r$).
For our three interferometer configurations, the values of $\mathcal B$ are 
\begin{equation}
\mathcal B_\text{WB} = 15\;, \quad
\mathcal B_{500} = 33\;, \quad
\mathcal B_{1000} = 66\;.
\label{calBvalues}
\end{equation}

When both ITM and SRM are tilted through the same angle $\theta$ about uncorrelated axes, the total power transferred into parasitic modes is twice
as large as Eq.\ (\ref{deltatau})
[i.e., $\delta_t$ is twice (\ref{deltatau})],
and correspondingly the
fractional increase of gravitational wave noise (due to both loss of signal power and increase
of shot noise) is
\begin{equation}
\left({\delta S_h\over S_h}\right)_\text{transmission} =
\delta_t = \mathcal B^2 k^2 \langle r^2\rangle \theta^2 \;.
\label{dShTransmission}
\end{equation}
Inserting the above values for $B$ and $\langle r^2\rangle$ and insisting
that 
$S_h$
not increase by more than one per cent, we obtain
the following constraints on the ITM and SRM tilt angles:
\begin{eqnarray}
\theta_{1\%}^\text{MH\;WB} &=& 1.6\times 10^{-8}\;, \quad 
\theta_{1\%}^\text{sph\;WB} = 2.4\times 10^{-8}\;,\nonumber\\
\theta_{1\%}^\text{MH\;500} &\agt& 0.7\times 10^{-8}\;,\quad
\theta_{1\%}^\text{sph\;500} \agt1.1\times 10^{-8}\;  \nonumber\\
\theta_{1\%}^\text{MH\;1000} &\agt& 0.4\times 10^{-8}\;; \quad
\theta_{1\%}^\text{sph\;1000} \agt0.6\times 10^{-8}\;.\nonumber\\
\label{thetaTransmission}
\end{eqnarray}

For the narrowbanded interferometers, the failure of the geometric
optics limit dictates that these estimates of $\theta_{1\%}$ are lower
limits; hence the ``$\agt$''.  As in the case of signal light in an arm 
cavity reflecting off the SR cavity, so also here, we \emph{guess} that 
these estimates are fairly good for narrow banding at 500 Hz and are
roughly a factor 2 too severe at 1000 Hz, so
\begin{eqnarray}
\theta_{1\%}^\text{MH\;500} &\simeq& 0.7\times 10^{-8}\;,\quad
\theta_{1\%}^\text{sph\;500} \simeq1.1\times 10^{-8}\;  \nonumber\\
\theta_{1\%}^\text{MH\;1000} &\simeq& 0.7\times 10^{-8}\;; \quad
\theta_{1\%}^\text{sph\;1000} \simeq1.1\times 10^{-8}\;.\nonumber\\
\label{thetaTransmission1}
\end{eqnarray}

Equations (\ref{thetaTransmission}) for wideband interferometers
and (\ref{thetaTransmission1}) for narrowband 
[and (\ref{thetaSignal1}) for $\theta_{1\%}^\text{MH\;1000}$]
are the most severe of all our tilt constraints.

\subsubsection{Reference light for baseline readout}

The baseline design for the advanced LIGO interferometers includes an output
mode cleaner and homodyne readout.  As we have seen [Eq.\ (\ref{ShShot})], 
the mode cleaner makes 
$S_h$
insensitive to (first-order) losses of
reference power into parasitic modes.

\subsubsection{Transmission of RF reference light through a power recycling cavity
without an output mode cleaner}
\label{sec:ReferenceTilt}

In LIGO-I interferometers, by contrast with the baseline design of 
advanced LIGO, there is no output mode cleaner, and heterodyne readout
is used in place of homodyne; i.e., the reference light is radio-frequency (RF)
sidebands, transmitted 
to the output port via the RF cavity (which is the same as the power recycling cavity 
since LIGO-I has no SR mirror).
In this section we shall compare our approximate 
analysis
with more careful analyses of LIGO-I, by analyzing this type of readout.

Suppose that the PRM or ITM is tilted through an angle $\theta$ and 
is thereby given the
space-dependent displacement $\delta z = \theta r \sin\varphi $.  Then,
in the geometric optics limit, the RF reference light acquires, when passing
through the PR cavity, a space-dependent phase shift 
$(\mathcal F / \pi) k
2\delta z$,\footnote{In 
\cite{DOSTVshortdoc} this equation was missing the factor 2, which
produced factor 2 errors in the resulting constraints on tilt; cf.\ Sec.\ IV$,$E$,$5 of \cite{DOSTVshortdoc}.
} 
where $\mathcal F$ is the cavity finesse.  The reference light
emerging from the cavity therefore has the form $u'_r = u_0 e^{i (\mathcal F
/\pi) k \delta z}$, for which the fraction of light power in parasitic modes is
\begin{equation}
 \left\langle u_0, \left(2{\mathcal F\over\pi}\right)^2 k^2 \delta z^2 \;u_0\right\rangle
= {1\over 2} \left( {2 \mathcal F\over \pi } k\right)^2 \langle r^2\rangle \theta^2\;.
\label{deltar}
\end{equation}
When both the PRM and the ITM are tilted through the same angle $\theta$
but around uncorrelated axes, the parasitic mode power is twice as large
[so $\delta_r$ is twice (\ref{deltar})],
and the fractional 
decrease of signal power, and corresponding increase of gravity-wave noise, due to the loss of this reference-light power, are then
\begin{equation}
\left({\delta S_h\over S_h}\right)_\text{reference} =
\delta_r =  \left( {2\mathcal F\over \pi} k\right)^2 \langle r^2\rangle \theta^2\;.
\label{dShRef}
\end{equation}
For the LIGO-I interferometers $\mathcal F$ is rather large, 
$\mathcal F \simeq 105$, 
which produces
a strong sensitivity to mirror tilt.  By contrast, for advanced
LIGO $\mathcal F$ is  smaller, 
$\mathcal F \simeq 50$, 
which (as we shall see) compensates for the larger beam. The result  would be
about the same sensitivity to tilt as for LIGO-I, if there were no output mode cleaner in
advanced LIGO and
heterodyne readout were used.

Inserting the LIGO-I finesse 
$\mathcal F \simeq 105$ 
and  mean square beam radius\footnote{LIGO-I 
has asymmetric arm cavities; the  2.9 cm that we use
is the root mean square of the beam radii on the ITM and ETM. 
}
$\langle r^2\rangle = b^2 = (2.9 \text{cm})^2$,
and constraining the 
increase of $S_h$
to less than one per cent, we obtain the following constraint
on the ITM and PRM tilts:
\begin{equation}
\theta_{1\%}^\text{LIGO-I} \simeq 0.9 \times 10^{-8}\;.
\end{equation}
This differs by a factor $\simeq 1.5$ from the result of a 
much more careful computation
by Fritschel et.\ al.\ \cite{fritschel}, which gave\footnote{Their result (end of
Sec.\ 2.A) is $\theta < 1.0 \times 10^{-8}$ for the tilts in pitch and in yaw, 
corresponding to a constraint 
$\theta<\sqrt{2} \times 10^{-8} \simeq 1.4\times 10^{-8}$
on the magnitude of the vectorial tilt, for an 0.005 fractional decrease in amplitude
signal to noise, which corresponds to our one per cent increase in $S_h$.}
$\theta_{1\%}^\text{LIGO-I} = 1.4 \times 10^{-8}$.

If advanced LIGO were to use an RF readout and no mode cleaner, the RF
cavity would be bounded by the PR, SR, and ITM mirrors (with each round
trip of light encountering the ITM twice).  Uncorrelated tilts of all three
three mirrors would lead to $\delta S_h/S_h = \delta_r$ twice as
large as expression (\ref{dShRef}).  
Inserting the advanced LIGO finesse $\mathcal F \simeq 50$
and mean-square beam radius $\langle r^2\rangle = (6.95\text{cm})^2$
(fiducial mesa) 
and $(4.70 \text{cm})^2$ 
(fiducial 
Gaussian), and constraining the increase in 
$S_h$
to no more than one per cent, we would then obtain 
\begin{equation}
\theta_{1\%}^\text{MH} \simeq0.4\times 10^{-8}\;, \quad 
\theta_{1\%}^\text{sph} \simeq 0.6\times 10^{-8}\;,\nonumber\\
\label{thetaReference}
\end{equation}
As we have seen, an output mode cleaner will remove this 
increase of $S_h$,
making these constraints no longer needed.

\subsection{Decrease in Signal Strength 
Due to Mirror Figure Errors}
\label{sec:FigureShot}

The loss of signal strength and resulting increase in $S_h$, caused by 
mirror figure errors,
is given by
the same equation $S_h \propto 1+2\delta_0 + 2 \delta_s + \delta_t + \{\delta_r
\hbox{ or } 0\}$ as for mirror tilt [Eq.\ (\ref{ShShot})], but now $\delta_\ell$ is the fraction of power in 
the parasitic components of mode $u'_\ell$ due to figure errors rather than tilt.

\subsubsection{Carrier light in arm cavity}

Deformations of the ITM and ETM, with optimized tilt compensation, 
drive a fraction $\delta_0 = \beta_{1cE}^2 + \beta_{1cI}^2$
into parasitic modes [Eqs.\ (\ref{uDeformedCompensated}) and (\ref{beta1c})].  Assuming the
same peak-to-valley deformations $\Delta z$ in the two mirrors' central regions, we obtain for the 
fractional increase in 
gravitational-wave noise
\begin{equation}
\left( {\delta S_h \over S_h} \right)^\text{MH} = 2\delta_0 =
4 \beta_{1c}^2 \simeq 0.01 \left( {\Delta z}
\over 15\text{nm}\right)^2\;.
\label{dShCarrierDeform}
\end{equation}
Correspondingly, to keep the noise
increase below one per cent, we must constrain the ITM
and ETM deformations to 
\begin{equation}
\Delta z_{1\%}^\text{MH} \simeq 15 \text{nm}
\label{DzCarrier}
\end{equation}
We have not carried out an analysis of the influence of the ITM and SRM
mirror deformations on the 
fiducial
Gaussian arm cavity modes, and so cannot say what the analogous constraint is in the 
Gaussian
case.

\subsubsection{Signal light in arm cavity}

As for tilt, so also for figure errors, the SR cavity amplifies the influence of the errors
$\Delta z$ by a factor $\mathcal A$ [Eq.\ (\ref{calA})], so the ITM and SRM deformations 
move a fraction $\delta_s = {\mathcal A}^2 (\beta_{1c\;I}^2 + \beta_{1c\;SR}^2)$ of the
arm cavity's signal light into parasitic modes.  When the two figure errors have the same
magnitude and we wish to keep the resulting 
loss of signal power and increase of $S_h$
below one per cent, this
gives rise to constraints on the ITM and SRM figure errors that are $1/\mathcal A$ more severe
than (\ref{DzCarrier}). Using the values (\ref{calAvalues}) of $\mathcal A$
for our three interferometers (wide-band, narrowbanded at 500 Hz and narrowbanded at 1000 Hz),
and increasing the limit for the 1000-Hz narrowbanded case by a factor 2
due to failure of the
geometric-optics limit [cf.\ Eq.\ (\ref{thetaSignal1}) and associated discussion], 
we obtain the constraints
\begin{eqnarray}
\Delta z_{1\%}^\text{MH\; WB} &\simeq& 55 \text{nm}\;, \nonumber\\
\Delta z_{1\%}^\text{MH\; 500} &\agt& 10 \text{nm}\;, \hbox{ with a guess of } \simeq 10 \text{nm}\;, \nonumber\\
\Delta z_{1\%}^\text{MH\; 1000} &\agt& 2.6 \text{nm}\;, \hbox{ with a guess of } \simeq 5 \text{nm}\;. \nonumber\\
\label{DzSignal}
\end{eqnarray}

\subsubsection{Transmission of signal light through SR cavity}
\label{sec:TransmissionFigure}

By the same analysis as for mirror tilt (Sec.\ \ref{sec:TransmissionTilt}), deformations 
$\delta z(x,y)$ of the
ITM and SRM by the same peak-to-valley amounts $\Delta z$ produce an increase in 
$S_h$
given by
\begin{eqnarray}   
\left({\delta S_h\over S_h}\right)_\text{transmission} &=&
\delta_t = 2 \mathcal B^2 k^2 \langle (\delta z)^2\rangle  \nonumber\\
&=& {1\over 4} \mathcal B^2 k^2 (\Delta z)^2 \;
\label{dShTransmissionDeform}
\end{eqnarray}
[cf.\ Eq.\ (\ref{dShTransmission}) with $\langle( \delta z)^2 \rangle = \langle (\theta r \sin\varphi)^2
\rangle = {1\over 2}\langle r^2\rangle \theta^2$].
Here $\langle (\delta z)^2 \rangle$ is the mean square deviation of ITM or SRM
height from the desired figure, and we have approximated this by half the squared amplitude
of mirror height fluctuations, which is 1/8 the square of the peak to valley height fluctuations,
$(\Delta z)^2/8$.  Inserting the values of $\mathcal B$ for our three interferometers
[Eq.\ (\ref{calBvalues})] and requiring that 
$S_h$ 
increase by no more than one per cent, we
obtain the following constraints on the ITM and SRM peak to valley deformations:
\begin{eqnarray}
\Delta z^\text{MH\; WB}_{1\%} &=& \Delta z^\text{sph\; WB}_{1\%} \simeq 2 \text{nm},
\nonumber\\
\Delta z^\text{MH\; 500}_{1\%} &=& \Delta z^\text{sph\; 500}_{1\%} \alt 1 \text{nm}, \hbox{ with a guess} \simeq 1 \text{nm}\;,
\nonumber\\
\Delta z^\text{MH\; 1000}_{1\%} &=& \Delta z^\text{sph\; 1000}_{1\%} \alt 0.5 \text{nm}, \hbox{ with a guess} \simeq 1 \text{nm}\;. \nonumber\\
\label{DzTransmission}
\end{eqnarray}
Here as in Eqs.\ (\ref{DzSignal}), (\ref{thetaTransmission1})   and (\ref{thetaSignal1}), the breakdown of the geometric optics limit in the SR cavity has dictated a lower limit and a guess for the narrowbanded interferometers.

These are the most serious of our constraints on the mirror figures of
advanced interferometers, and they are the same for 
fiducial
mesa and 
fiducial 
Gaussian
beams, because transmission through the SR cavity is governed (at least roughly) by 
geometric optics.  The one small difference is that the central region over which
the peak-to-valley deformations are constrained (the region containing $\sim 95$ per cent of the light
power) is larger for 
the fiducial
MH mirrors
(central radius about 10 cm) than for the 
fiducial 
Gaussian mirrors (central
radius about 8 cm).

The mirror-figure constraint (\ref{DzTransmission}) is sufficiently severe, at
least in the case of narrowbanded interferometers, that it might be worth 
considering reducing the degeneracy of the SR cavity by making the entrance faces
of the ITM's into lenses that bring the signal light (and inevitably also the carrier
light) to a focus somewhere near the SRM (and PRM).  Since the  constraint
(\ref{DzTransmission}) is the same, whether the mirrors are MH or spherical,
this recommendation is not dependent on whether mesa beams are implemented.

\subsubsection{Reference light for baseline readout}

As for mirror tilt, so for figure errors, the output mode cleaner in advanced LIGO makes
$S_h$
insensitive to the loss of reference light  into parasitic modes.

\subsubsection{Transmission of RF reference light through power recycling
cavity without an output mode cleaner}

In the 
LIGO-I 
case of no output mode cleaner and heterodyne readout with RF sideband
light, the loss of reference light to parasites does increase 
$S_h$.
By the same argument as for mirror tilt (Sec.\ \ref{sec:ReferenceTilt}), the net 
increase of $S_h$
due to 
deformations $\delta z(x,y)$ of the ITM
and PRM is\footnote{In 
\cite{DOSTVshortdoc} the factor 2 in $(2{\mathcal F}k/\pi)^2$
was missing, which
produced factor 2 errors in the resulting constraints 
(\ref{DeltazRef})
on $\Delta z$; cf.\ Sec.\ IV$\,$F$\,$4 of \cite{DOSTVshortdoc}.
}
\begin{eqnarray}   
\left({\delta S_h \over S_h}\right)_\text{reference} &=&
\delta_r = 2\left( {2\mathcal F \over \pi} k\right)^2 \langle (\delta z)^2 \rangle \nonumber\\
&=& {1\over 4}\left( {2\mathcal F \over \pi} k\right)^2 (\Delta z)^2 \;
\label{dShRefrenceDeform}
\end{eqnarray}
[cf.\ Eq.\ (\ref{dShRef}) with $\langle( \delta z)^2 \rangle = \langle (\theta r \sin\varphi)^2
\rangle = {1\over 2}\langle r^2\rangle \theta^2$].
For advanced LIGO with RF readout and no mode cleaner, there is an extra factor 2
in Eq.\ (\ref{dShRefrenceDeform}) due to change of configuration of the RF cavity (the
added SRM and the light encountering the ITM twice on each round trip).
Inserting the finesses of LIGO-I 
($\mathcal F \simeq 105$) 
and advanced LIGO
($\mathcal F \simeq 50$) 
and insisting that 
$S_h$
not be increased by
more than one per cent, we obtain the following constraints on the central-region
peak-to-valley deformations of the PRM and SRM:
\begin{eqnarray}
\Delta z^\text{LIGO-I}_{1\%} &=&  0.5 \text{nm}\;,
\nonumber\\
\Delta z^\text{MH}_{1\%} &=& \Delta z^\text{sph}_{1\%} = 0.5 \text{nm}\;.
\label{DeltazRef}
\end{eqnarray}
These constraints are rather severe, but for advanced LIGO the constraint is 
actually irrelevant because of the output
mode cleaner (and the planned use of homodyne readout rather than heterodyne). 

\section{Conclusions and Future Research}
\label{sec:Future}

The thermoelastic benefits of MH mirrors are sufficiently great, and the tightened
constraints that they place on mirror figures, positions and tilts are sufficiently modest,
that MH mirrors have been adopted by the LSC as an option for advanced LIGO,
and will be incorporated into future modeling along with spherical mirrors.

Two methods of manufacturing MH mirrors are being explored by the LIGO Laboratory:  
diamond cutting, and evaporative coating.  The immediate goal is to determine the   accuracy and reproducibility
with which the desired MH mirror figures can be produced by each of these methods.  

If MH mirrors are to be considered seriously for advanced LIGO, it is necessary to
develop laboratory experience with them and with the mesa-beam optical cavities
that they produce.   An experimental effort in this direction has been initiated
by Phil Willems of the LIGO Laboratory at Caltech.

As we have discussed in Secs.\  \ref{sec:RecyclingCavities} and
\ref{sec:TransmissionFigure}, the constraints
on mirror figure in the recycling cavities are rather worrisome
for both spherical mirrors and MH mirrors.  
More accurate
studies of this are needed.  As one aspect of these studies, it is important to check
our geometric-optics-based claims that the 
fiducial
MH recycling cavities are not much
more sensitive to mirror figure errors than the 
fiducial 
Gaussian-beam
cavities.  

If the recycling-cavity-induced
constraints on mirror figure are found to be as serious as our estimates
suggest, then it seems worthwhile to carry out
studies of the option of converting the input faces of the ITM's
into lenses that make the recycling cavities much less degenerate.
PhD students in Thorne's research group
may carry out these studies as part of the broadening experience that is 
a standard part of their education.

Bill Kells has suggested the possibility
of operating the advanced LIGO interferometers initially with spherical mirrors and
Gaussian beams, and later switching to mesa beams by altering only the
ETM's.  It seems to us that
this option is worth detailed study.  If the ITM input faces are turned into
lenses that reduce the near degeneracy of the recycling cavities, then it might
be possible to keep the lenses weak enough that the figures of the recycling
mirrors can be the same for Gaussian and mesa beams, while still relaxing
the mirror figure constraints to an acceptable level.  This needs study.

\begin{acknowledgments}
For helpful discussions and advice we thank, among others, 
Raymond Beausoleil,
Alessandra Buonanno,
Vladimir Braginsky,
Yanbei Chen,
Peter Fritschel,
Greg Harry,
Bill Kells,
David Shoemaker,
Robert Spero,
Ken Strain, 
Rai Weiss,
Phil Willems,
Stan Whitcomb,
Farid Khalili,
Mike Zucker, 
and especially Garilynn Billingsley 
and Guido M\"uller.  
We thank Billingsley for providing us with the mirror figure map
(Fig.\ \ref{fig:GariMap}) used
in our study of the influence of mirror figure errors, 
we thank M\"uller for a detailed critique of the manuscript
(in his role as LSC referee) which resulted in significant changes,
and we
thank Patrice Hello, Jean-Yves Vinet, Brett Bochner, and others 
who developed and refined the FFT code that ED'A adapted for
use in her calculations of the influence of mirror tilt, displacement,
and figure errors.
This research was supported in part
by NSF grants  PHY-0098715 and PHY--0099568, 
by the Russian Foundation for Fundamental Research grants
\#96-02-16319a and \#97-02-0421g, and (for SPV) by the
NSF through Caltech's Institute for Quantum Information.
\end{acknowledgments}

\appendix

\section*{Appendix:  Approximate Formulae for Mesa Modes}

By inserting into Eq.\ (\ref{new}) the asymptotic expansion (at large agument) for the
Bessel function
\begin{equation}
I_0(z) ={ 1\over\sqrt{2\pi z}} e^z\;,
\label{asymptoticBessel}
\end{equation}
setting $b=1$,  $r_o = D-\xi$, and 
\begin{equation}
\alpha = \sqrt{(1+i)/2}\;,
\label{alpha}
\end{equation}
we bring Eq.\ (\ref{new}) into the approximate form
\begin{equation}
U(D,r) =\sqrt{\pi\over \alpha^2 r}\int_0^D \sqrt{D-\xi}\exp[{-\alpha^2(r-D+\xi)^2} ]d\xi\;.
\label{approximate0}
\end{equation}
By expanding $\sqrt{D-\xi}$ as a power series in $\xi/D$ up to some order $n$,
and then performing the integral in Eq.\ (\ref{approximate0}) 
analytically, we obtain expressions
for $U(D,r)$ with various accuracies.  The least accurate expression, $n=0$ (obtained
by setting $\sqrt{D-\xi} = \sqrt{D}$, integrating, and discarding a term proportional
to $\text{erfc}(\alpha r)$ that is negligible compared to $\text{erfc}[\alpha(r-D)]$ at the
relevant radii, $r \sim D$ or larger)  is 
\begin{equation}
U_0(D,R) = {\pi \over 2\alpha^2}\sqrt{ D\over r} \text{erfc}[\alpha(r-D)]\;.
\label{U0}
\end{equation}
Here erfc$(z)$ is the complementary error function, $1-\text{erf}(z)$.
At order $n=3$ we get a much more accurate expresssion:
\begin{eqnarray}
&&U_3(D,r) = {1\over 64\alpha^5 D^2 \sqrt{Dr}} \nonumber\\
&&\times  \left\{ \pi \left[\alpha^3(2r^3-10Dr^2 + 30D^2r+10D^3) +\alpha(3r-5D)\right]\right.
\nonumber\\
&&\quad \times \Big[ \text{erfc}[\alpha(r-D)] - \text{erfc}[\alpha r]\Big] \nonumber\\
&&  +2\sqrt{\pi} \left[ e^{-\alpha^2r^2}[\alpha^2(15D^2-5Dr+r^2)+1]\right. \nonumber\\
&&\quad \left.\left.
- e^{-\alpha^2(r-D)^2}[\alpha^2(11D^2-4Dr+r^2)+1] \right] \right\}\;.
\label{U3}
\end{eqnarray}
Here erfc is the complementary error function, erfc$(z) = 1- \text{erf}(z)$.

\begin{figure}
\begin{center}
\includegraphics[width=3.4in]{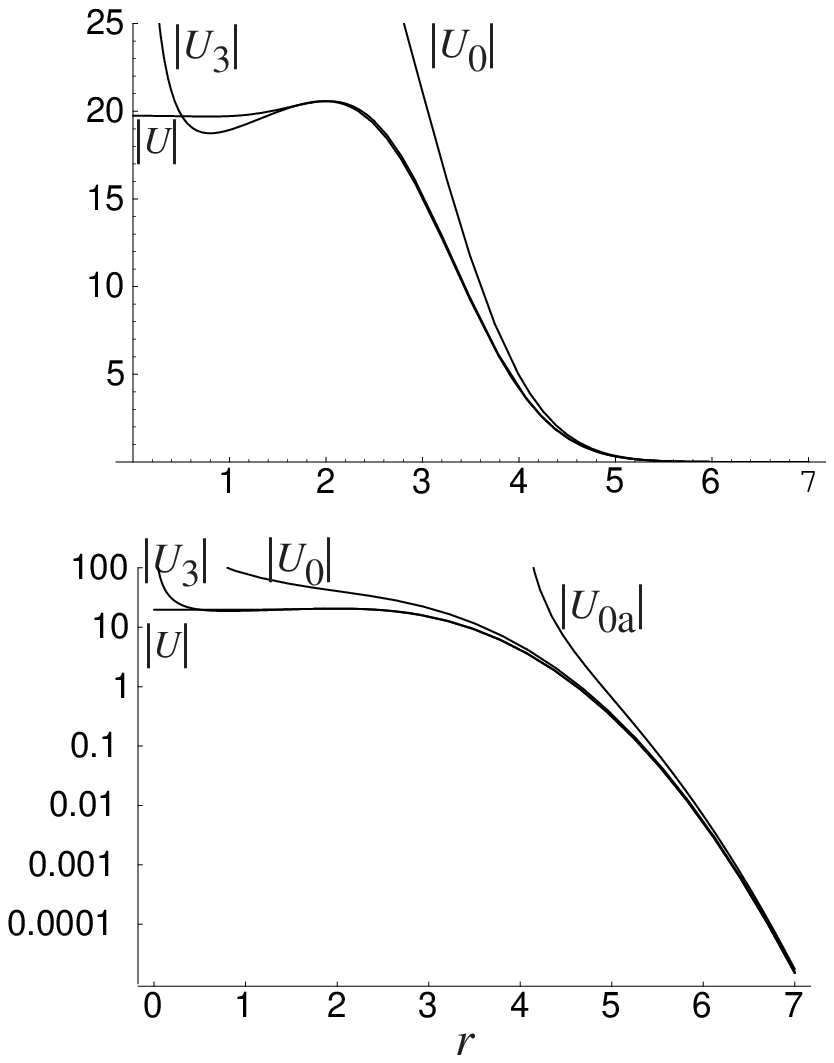}
\end{center}
\caption{Comparison of the moduli of the exact flat-topped mode $U(4,r)$, the zero-order
approximation  $U_0(4,r)$ [Eq.\ \ref{U0})],  the third-order approximation
$U_3(4,r)$ [Eq.\  (\ref{U3})], and the asymptotic approximation to the zero-order
approximation, $U_{0a}(4,r)$ [Eq.\ (\ref{absU0a})].  Top: linear plot; bottom: logarithmic plot.
}
\label{ModeGraphs}
\end{figure} 

The two approximations $U_0(4,r)$ and $U_3(4,r)$ are compared with the exact
mode $U(4,r)$ in Fig.\  \ref{ModeGraphs}.  At $r>2$, $U_3$ is highly accurate;
at $r>5$, $U_0$ is highly accurate.  The analytic formula (\ref{U0}) for $U_0$ shows
that the diffraction-induced tail of this mode falls off very rapidly outside $r=D$ ---
slightly more rapidly than would the tail of a Gaussian centered on $r=D$.  Of course,
this is to be expected since $U(D,r)$ is constructed from a superposition of Gaussians that are centered on radii $\le D$.

A useful but cruder approximation to $U(D,r)$ can be obtained by inserting the asymptotic expansion of $\text{erfc}[\alpha(r-D)]$ into Eq.\ (\ref{U0}) for $U_0$.  The result is
\begin{equation}
U_{0a}(D,r) ={ \sqrt{\pi D/r}\over 2 \alpha^3 (r-D)} e^{-\alpha^2(r-D)^2}\;,
\label{U0a}
\end{equation}
which has a modulus
\begin{equation}
|U_{0a}(D,r)| = \sqrt{\pi D\over \sqrt2 r} {e^{-(r-D)^2 /2} \over r-D}\;.
\label{absU0a}
\end{equation}
This modulus is accurate to within several tens of per cent in the regime of interest for 
diffraction, $D\sim 4$ and $R \agt 6$ (in units of $b$).  By inserting this approximation
into Eqs.\ (\ref{calLclip}) and (\ref{uMH}) for diffraction losses in the clipping approximation, we deduce that
the diffraction losses for mesa beams scale with beam-spot radius $D$ as 
\begin{equation}
\mathcal L_\text{clip}^\text{mesa} \propto \exp({-2RD/b^2})\;.
\label{Lclipmesa}
\end{equation}
Here $R$ is the mirror radius and we have restored the diffraction lengthscale 
$b$, which was set to one throughout this appendix.

\end{document}